\documentclass[a4paper,aps,pra,twocolumn,amsmath,amssymb]{revtex4}

\usepackage{epsfig}
\usepackage{amsmath}
\usepackage{graphicx}
\usepackage{bm}

\begin{document}

\title{Quantum Optomechanics - throwing a glance}\thanks{This work was published in J.\ Opt.\ Soc.\ Am.\ B \textbf{27}, A189-A197 (2010).}
\author{Markus Aspelmeyer,$^{1,}$\footnotemark\footnotetext{email:\ markus.aspelmeyer@quantum.at} Simon Gr\"oblacher,$^{1}$ Klemens Hammerer,$^{2,3}$ and Nikolai Kiesel$^{1}$}
\affiliation{
$^1$ Faculty of Physics, University of Vienna, Boltzmanngasse 5, A-1090 Vienna, Austria\\
$^2$ Institute for Quantum Optics and Quantum Information (IQOQI), Austrian Academy of Sciences, Technikerstra\ss e 21a, A-6020 Innsbruck, Austria\\
$^3$ Institute for Theoretical Physics, University of Innsbruck, Technikerstra\ss e 25, A-6020 Innsbruck, Austria}

\begin{abstract}
Mechanical resonators are gradually becoming available as new quantum systems. Quantum optics in combination with optomechanical interactions (quantum optomechanics) provides a particularly helpful toolbox for generating and controlling mechanical quantum states. We highlight some of the current challenges in the field by discussing two of our recent experiments.
\end{abstract}

\maketitle

\begin{center}
\large\textbf{Introduction}\\
\end{center}

Today's quantum science has achieved an amazing level of control over individual quantum states. They are implemented in a large variety of physical systems from photons, electrons, or atoms up to solid-state systems such as quantum dots or even electronic circuits~\cite{Southwell2008,Osborne2008}. This advancement has led to both more and more realistic schemes for practical quantum information processing ~\cite{Zoller2005} and to a “quantum renaissance” for fundamental experiments~\cite{Aspelmeyer2008b} by allowing the access to hitherto unachieved parameter and resolution regimes for novel tests of quantum theory.

Mechanical resonators are playing an increasing role in theoretical and experimental quantum physics, and they are now at the verge of becoming a new entry in the zoology of "tamed" quantum systems. In contrast to atomic systems, where quantum control of the mechanical motion is state of the art~\cite{Blatt2008,Jost2009}, mechanical resonators span the size range from hundreds of nanometers in the case of nano-electro-mechanical or nano-opto-mechanical systems (NEMS/NOMS) to tens of centimeters in the case of gravitational wave antennae. The potential for quantum-nanomechanical devices has already been envisioned in the 1990s~\cite{Cleland1996,Cleland1998,Schwab2005}, and the successive developments in micro- and nanomechanical engineering eventually triggered the recent enormous experimental progress towards achieving this enticing goal. Within a very short time scale of only a few years, research on mechanical systems has generated a new interdisciplinary community of scientists who seek to achieve control over mechanical quantum states~\cite{Cho2010}. A broad variety of experimental approaches has emerged including the coupling of mechanical systems to single electrons via spin~\cite{Rugar2004} or charge~\cite{LaHaye2004}, to Cooper pairs via microwave cavities ~\cite{LaHaye2009}, to phase qubits via charge~\cite{OConnell2010}, or to photons inside an optical cavity~\cite{Kippenberg2005,Gigan2006,Arcizet2006b,Thompson2008,Regal2008}. Only recently, coherent single-phonon control in a high-frequency micromechanical resonator has been demonstrated~\cite{OConnell2010}.

Quantum optics provides a well-developed toolbox for generating, controlling and manipulating quantum states of a mechanical system, and a particularly strong analogy exists for the optomechanical case: in essence, the well-known situation of two-mode quantum optics, in which two optical modes interact via a nonlinear medium, can be mapped onto a quantum-optomechanical situation, in which an optical mode and a mechanical mode interact via the transfer of photon momentum, i.e., via radiation pressure inside an optical cavity. This analogy has been pointed out from very early on and is based on two remarkable facts: first that the single-mode quantum description of an optical field is essentially equivalent to that of a harmonic oscillator, and second that the interaction between light and a mechanical oscillator can resemble a nonlinear (Kerr-type) interaction if the light is confined in a cavity that is modified by the mechanical motion. The latter may be implemented for example via direct changes in the cavity length~\cite{Law1994}, by dispersion~\cite{Thompson2008}, or by optical near-field effects~\cite{Li2008,Eichenfield2009,Roels2009,Anetsberger2009}.

\begin{figure*}[htbp]
\centerline{\includegraphics[width=0.67\textwidth]{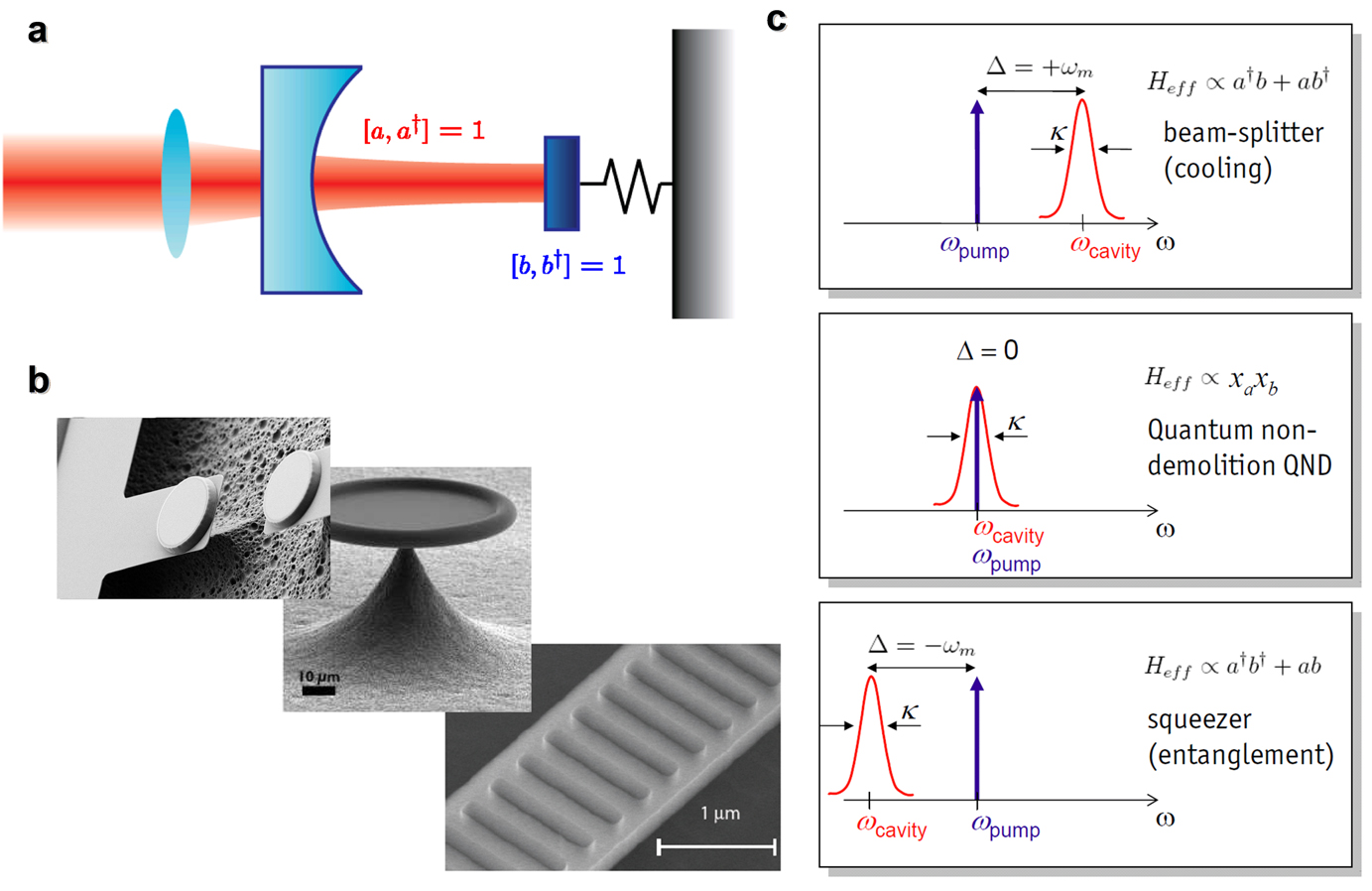}}
\caption{\textbf{a} Schematic representation of an optomechanical system. The cavity mode is an optical harmonic oscillator (described by creation and annihilation operators $a^{\dag}$ and $a$ of the optical field) that is coupled to a mechanical harmonic oscillator (described by creation and annihilation operators $b^{\dag}$ and $b$ of the mechanical motion), e.g., via the mechanical modulation of the cavity length. \textbf{b} Recent examples of mechanical resonator designs in a Fabry-P\'{e}rot cavity~[66], a microtoroid structure~[59], and an optomechanical crystal~[23]. \textbf{c} The choice of detuning $\Delta$ of the driving laser frequency $\omega_{pump}$ with respect to the cavity resonance frequency $\omega_{cavity}$ allows to engineer the optomechanical interaction and provides access to the toolbox of two-mode quantum optics. Red detuning (blue detuning) gives rise to a beamsplitter (two-mode squeezing) interaction, which is of relevance, e.g., for optical cooling or quantum state transfer (for the generation of optomechanical entanglement). Resonant driving ($\Delta=0$) allows for example for quantum nondemolition measurements.} \label{fig1}
\end{figure*}

The idea that mechanical properties can be modified by radiation pressure forces goes back to the pioneering works of Braginsky~\cite{Braginsky1970}. The first papers that made use of the quantum optics analogy and explicitly discussed the generation of quantum effects through radiation pressure inside an optical cavity appeared in the 1990's and suggested among other ideas the generation of squeezed light via the optomechanical Kerr-nonlinearity~\cite{Fabre1994,Mancini1994}, optomechanical schemes for photon number quantum non-demolition measurements~\cite{Milburn1994,Pinard1995}, mechanical quantum noise reduction, i.e., cold damping, via feedback techniques~\cite{Mancini1998,Cohadon1999} or the generation of nonclassical states of mechanical systems by optomechanical interactions~\cite{Bose1997,Mancini1997}. At that time, however, experiments were not advanced enough to produce the required optomechanical interaction strengths. Only one pioneering experiment in 1983 had previously demonstrated radiation-pressure based nonlinearities in an optical cavity~\cite{Dorsel1983}. In 1999 the principle of feedback cooling by radiation pressure of a mechanical mirror mode has been demonstrated~\cite{Cohadon1999}, and in 2004 a first analogue to laser cooling was shown for a micromechanical resonator using photothermal forces~\cite{Metzger2004}. Since then a plethora of experiments and theory proposals have successfully established quantum optomechanics as a viable approach to achieve and control the quantum regime of mechanical resonators~\cite{Kippenberg2008,Aspelmeyer2008,Favero2009,Marquardt2009,Genes2009}. Recently, even ultracold atoms have emerged as ideal test systems for cavity optomechanical interactions by using a collective mechanical degree of freedom of the atomic cloud, which is strongly coupled to an optical cavity field~\cite{Murch2008,Brennecke2008}. With the availability of high-quality optomechanical devices, these early theories have now become an important basis for a whole new field of quantum optomechanics that aims at exploiting the quantum regime of mechanical resonators by means of quantum optics.

\begin{figure*}[htbp]
\centerline{\includegraphics[width=0.67\textwidth]{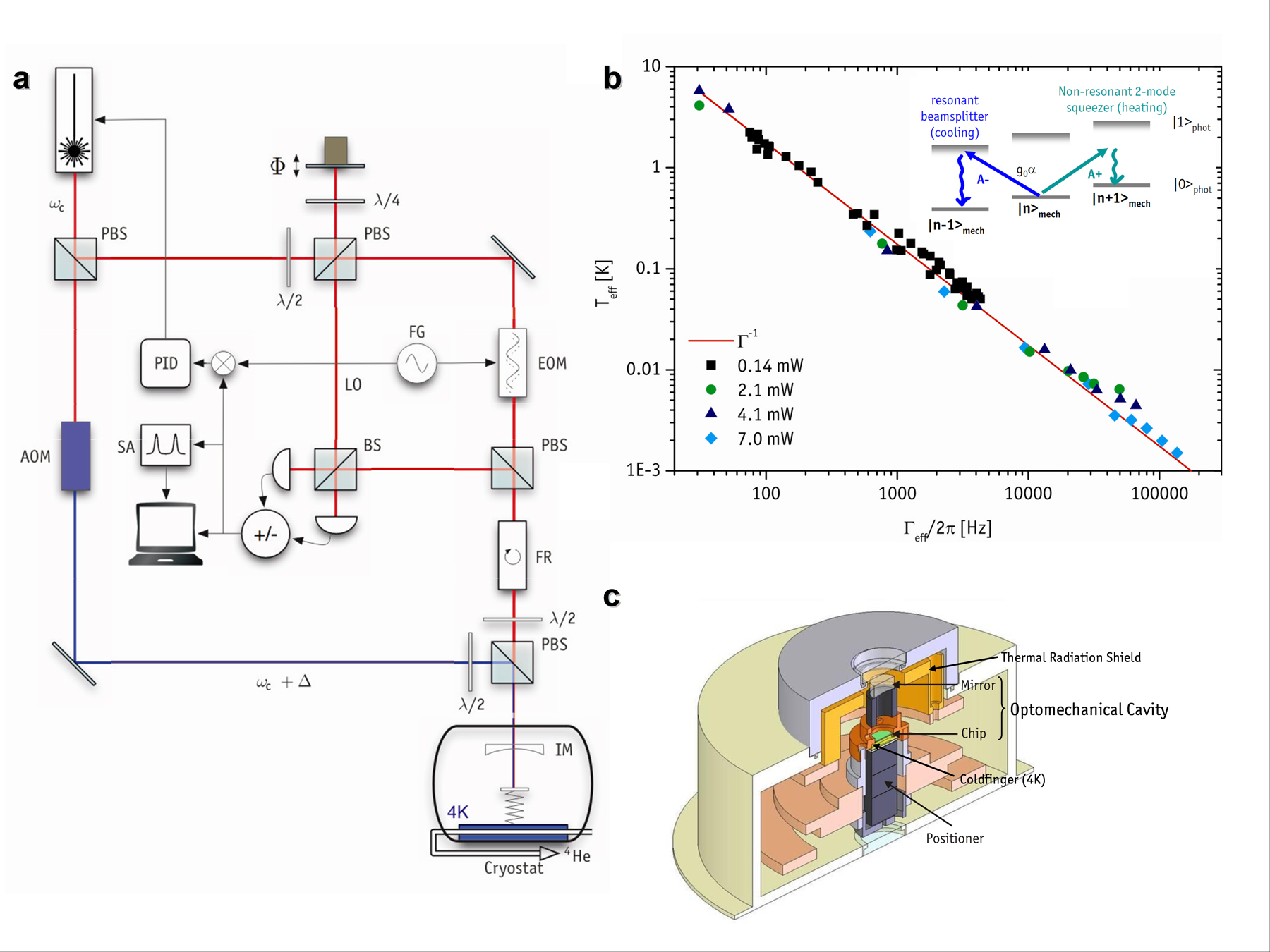}}
\caption{Mechanical Laser Cooling. \textbf{a} Sketch of the setup that was used to perform resolved sideband cooling of a micromechanical resonator in a cryogenic Fabry-P\'{e}rot cavity (from~[66]). The cavity is pumped by two laser fields at orthogonal polarizations, derived from an ultralow phase-noise laser source (at 1064~nm). A weak, resonant beam (red) is used to lock the laser to the cavity resonance frequency $\omega_c$. Its phase quadrature is modulated by the mechanical motion, which can be read out via homodyne detection of the reflected resonant pump field. This allows, e.g., to infer the effective temperature of the mechanical mode from the noise-power spectrum of the cavity-field phase quadrature. A second, stronger beam is red detuned by the mechanical resonator frequency ($\Delta=\omega_m$) and is used to cool the center of mass motion of this resonator mode. The cooling process (inset in \textbf{b}) takes place because the rate ($A-$) at which photons are scattered into the anti-Stokes sideband at the cavity resonance under extraction of a phonon is dominant compared to heating by scattering into the off-resonant Stokes sideband ($A+$). \textbf{b} The plot shows the calibrated effective mechanical mode temperature $T_{eff}$ versus the observed mechanical damping $\Gamma_{eff}$ for various power and detuning values of the cooling beam. No deviations from the theoretically expected power-law dependence (red solid line) can be observed, which demonstrates the absence of residual heating effects. In this experiment the fundamental mode of the micromechanical resonator has been cooled down to a mean occupation of $\langle n\rangle=32$ phonons, only limited by the unavoidable thermal coupling of the resonator to its environment. \textbf{c} Cross section of the $^4$He cryostat to precool the optomechanical system. The optical access is provided through a thermal radiation shield. The bulk input coupler is mounted inside the cryostat. The chip (green), with the mechanical resonator ($\omega_m\sim 2\pi\times 950$~kHz, $Q\sim 30000$) is cooled to 5~K. It can be aligned with respect to the second mirror forming the Fabry-P\'{e}rot cavity ($F=3900$) with a bandwidth of $2\pi\times 770$~kHz, allowing operation in the moderately resolved sideband regime.} \label{fig2}
\end{figure*}

The prototypical setup works as follows. Let us consider the case of a single-sided Fabry-P\'{e}rot cavity of length $L$, frequency $\omega_c$, and linewidth $\kappa$ with a suspended end mirror of mass $m$ and mechanical resonance frequency $\omega_m$ (see Fig.\ 1a). This configuration covers the essential features of quantum optomechanics and can be transferred easily to all other optomechanical geometries. For more detailed descriptions of the variety of cavity optomechanics experiments we refer the reader to some of the recent excellent overviews~\cite{Kippenberg2008,Aspelmeyer2008,Favero2009,Marquardt2009,Genes2009}. The fundamental optomechanical radiation-pressure interaction $H_{int}=-\hbar g_0 n_c X_m$ couples the photon number $n_c$ of the optical intracavity mode to the (dimensionless) position $X_m$ of the
mechanical mode ($\hbar$: Planck's constant), with the single photon coupling strength $g_0$. In the following, we focus on the situation in which the optomechanical
cavity is driven by a strong external pump laser at frequency $\omega_L$. As shown in~\cite{Wilson-Rae2007,Marquardt2007,Genes2008}, if the steady-state amplitude $\alpha_c$ of the cavity field is large, i.e., $|\alpha_c|\gg 1$, the drivingr esults in a linearized interaction described by
\begin{equation}
H_{int}=\hbar gX_cX_m=\hbar g(a_cb_m+a_c^{\dag}b_m^{\dag})+\hbar g(a_cb_m^{\dag}+a_c^{\dag}b_m)
\end{equation}
with an opto-mechanical coupling rate $g=g_0\alpha_c\sqrt{2}=\frac{2\omega_c}{L}\sqrt{\frac{P\kappa}{m\omega_m\omega_L(\kappa^2+\Delta^2)}}$  (following~\cite{Genes2008}) for an input power $P$ of the driving laser and effective spectral detuning $\Delta=\omega_c-\omega_L-\frac{g_0^2|\alpha_c|^2}{\omega_m}$ between the cavity and the pump laser. The full Hamiltonian of the system is $H=H_0+H_{int}$ with the free Hamiltonian $H_0=\hbar\Delta(a_c^{\dag}a_c)+\hbar\omega_m(b_m^{\dag}b_m)$. Equation (1) comprises two well-known interactions from two-mode quantum optics~\cite{Zhang2003}: the first term describes the transfer of energy between the driving laser and the joint excitations of the optical and the mechanical mode -- in quantum optics this interaction is known as two-mode squeezing or down-conversion~\cite{Leonhardt1997}. It allows to generate both correlations and anticorrelations between pairs of quadratures of the involved modes and, for the case of sufficiently strong interaction, it will produce Einstein-Podolsky-Rosen (EPR) entanglement between the optical and mechanical mode~\cite{Vitali2007}, fully analogous to the seminal demonstrations with two optical beams~\cite{Wu1987}. The second term describes energy transfer between the two modes and is equivalent to a quantum optical beamsplitter interaction. In optical experiments it is, e.g., at the heart of quantum state transfer protocols~\cite{Zhang2003,Parkins1999,Lukin2000,Julsgaard2004}. In the optomechanical case, as long as photons can leave the cavity, this interaction can be used to actually cool the mechanical mode (see below). It is an essential feature that we can now "tune in" the type of interaction by simply choosing the right detuning $\Delta$ between pump laser and optical cavity (see Fig.\ 1c). Let us consider the three most relevant cases:

For resonant driving ($\Delta=0$) both terms contribute equally and the interaction resembles a quantum-nondemolition (QND) interaction (see, e.g.,~\cite{Grangier1998}). For example, this is the basis of high-sensitivity position measurements of the mechanical oscillator, and of proposals to perform a nondemolition measurement of the intensity of an optical beam passing through an optomechanical cavity~\cite{Milburn1994,Pinard1995}. In essence, length changes that are induced by optomechanical coupling of different photon numbers do not affect the intensity response of the cavity, i.e., the light intensity remains unaltered. However, they leave a detectable signature in the cavity's phase response. In that way one can monitor the photon number without disturbing it. Some first experiments in this direction have been recently reported~\cite{Verlot2009}. Another interesting application of this scenario is the generation of optical squeezing (see references above).

For detuned driving at $\Delta=-\omega_m$ ($+\omega_m$) the first (second) term in Eq.\ (1) becomes resonant. For weak coupling ($g<\kappa$) the strength of the nonresonant terms is suppressed by the factor ($\kappa/\omega_m$)~\cite{Schliesser2006}; therefore, operation in the so-called resolved sideband regime, i.e., $\kappa\gg\omega_m$, is necessary for separating the two interactions. For increasing coupling strength the rotating-wave approximation will break down and one will have to take into account the full dynamics of Eq.\ (1).

While the analogy between the quantum optical and the optomechanical situation is striking, there are some obvious differences between a mechanical resonator and a photon. Because the mechanical frequencies involved, $\omega_m$, are typically well below $k_BT /\hbar$ at room temperature, we will always find a room-temperature mechanical resonator in a thermal, i.e., maximally mixed, state. In contrast, for optical frequencies $\omega_0\gg k_BT /\hbar$ for all relevant temperature scales; hence optical modes can, in the absence of noise, always be considered to be in a pure state. Sufficient purity of the mechanical mode will only be obtained for mode temperatures $T<\hbar\omega_m /k_B$, i.e., $T<50$~$\mu$K ($50$~mK) for a mechanical mode at $\omega_m/2\pi=1$~MHz ($1$~GHz). In addition, the interaction strength needs to be sufficiently large to overcome the unavoidable coupling of the optomechanical system to its environment. In the following we will describe two concrete experimental examples, mechanical laser cooling and strong optomechanical coupling, to discuss the current status of these challenges and to provide an outlook for future directions.\\

\begin{center}
\large\textbf{Optomechanical laser cooling}\\
\end{center}

As discussed before, for detuned driving at $\Delta=+\omega_m$ the beamsplitter interaction is dominant and energy is converted between the intracavity field and the resonator. Because the scattered photons leak out of the cavity, energy is dissipated from the optomechanical system. Several proof-of-concept experiments have demonstrated this radiation-pressure-based cooling effect~\cite{Gigan2006,Arcizet2006b,Thompson2008,Regal2008,Schliesser2006,Corbitt2007,Teufel2008,Wilson2009,Schliesser2008}; it has been shown theoretically that, in the sideband-resolved regime~\cite{Schliesser2008}, ground-state cooling of the mechanical resonator is possible in principle~\cite{Wilson-Rae2007,Marquardt2007,Genes2008}. To put it in the language of quantum-optomechanics: in this regime the resonant cooling (beamsplitter) interaction can become dominant over the nonresonant heating (two-mode squeezing) interaction and hence suppresses additional heating. This is reminiscent of sideband-resolved cooling of ions~\cite{Leibfried2003}, and indeed a close analogy exists~\cite{Wilson-Rae2007}. Interestingly, and in contrast to ion cooling, in the optomechanical case there is no resonant heating term due to the driving field and therefore mechanical laser cooling is intrinsically faster than cooling of ions. The final occupation is given by $\langle n\rangle\sim\kappa^2/4\omega_m^2$ and is equivalent to the Doppler temperature in atomic laser cooling~\cite{Stenholm1986}.

Yet, this cooling mechanism has to compete against additional sources of heating. Most importantly, thermal coupling of the mechanical resonator to its environment at temperature $T$ results in an additional heating rate $\Gamma_{th}=(k_BT)/(\hbar Q)$. Minimizing this effect requires either operation of an optomechanical cavity at low temperatures, i.e., in a cryogenic environment~\cite{Regal2008,Tittonen1999,Groeblacher2008}, or mechanical resonators with small dissipation, i.e., high quality factor $Q$ or, ideally, both. Second, phase-noise of the driving laser can give rise to an additional effective heat bath~\cite{Diosi2008}. It has recently been shown, however, that currently available laser systems are sufficiently stable to allow for ground-state cooling~\cite{Rabl2009b}. Finally, absorption of photons in the mechanical resonator will certainly lead to additional heating and has to be prevented by a proper choice of materials and lasers.

A series of recent experiments has attempted to combine these requirements~\cite{Teufel2008,Groeblacher2008,Groeblacher2009a,Schliesser2009,Rocheleau2010} and has demonstrated laser cooling of megahertz mechanical devices down to a level of 4 thermal quanta for nanoscale~\cite{Rocheleau2010} and of 30 and 60 thermal quanta for microscale resonators~\cite{Groeblacher2009a,Schliesser2009} (see Fig.\ 2 for a sketch of the experiment). In all cases, one of the above residual heating mechanisms eventually prevented ground-state cooling. In most cases~\cite{Teufel2008,Schliesser2009,Rocheleau2010}, absorption of cavity photons led to excess heating, while in~\cite{Groeblacher2008,Groeblacher2009a} the thermal coupling rate $\Gamma_{th}$ was still too large ($\Gamma_{th}\sim 1.4\times 10^7$) compared to the achieved optical cooling rate ($\Gamma_{bs}\sim 5\times 10^5$). For the latter case, improving the mechanical quality~\cite{Cole2008,Anetsberger2008} in combination with better cryogenic systems is an obvious option. In a conservative yet optimistic scenario, one could presently expect a mechanical device with $Q\sim 10^6$ operating inside a dilution refrigerator at $T\sim 100$~mK, which would result in a minimum excess heating rate of $\Gamma_{min}\sim 10^4$ and hence allow laser cooling into the ground state. Another strategy to minimize $\Gamma_{th}$ has recently been suggested: optical trapping of dielectric objects may lead to a significant improvement in mechanical isolation and could even allow optical ground-state cooling starting from room temperature~\cite{Romero-Isart2010,Chang2010,Barker2010}. In analogy to the ion case, readout of the ground state could be performed via sideband-spectroscopy of the pump field~\cite{Wilson-Rae2007}, where photons at the cavity frequency will be absent if the mechanical system is in its ground state, i.e., the mechanics cannot provide additional energy for the scattering process.

It should finally be noted that quantum optics provides also other means for cooling. The first proposal to achieve mechanical ground-state cooling has, for example, been made in the context of quantum atom optics~\cite{Wilson-Rae2004}]. In another approach, optical homodyning allows a direct readout of the mechanical phase space at the quantum limit. In combination with proper feedback, this monitoring can be used to actively drive the system into its quantum ground state~\cite{Mancini1998,Cohadon1999,Genes2008,Poggio2007,Kleckner2006b,Corbitt2007b}.\\

\begin{figure*}[htbp]
\centerline{\includegraphics[width=0.67\textwidth]{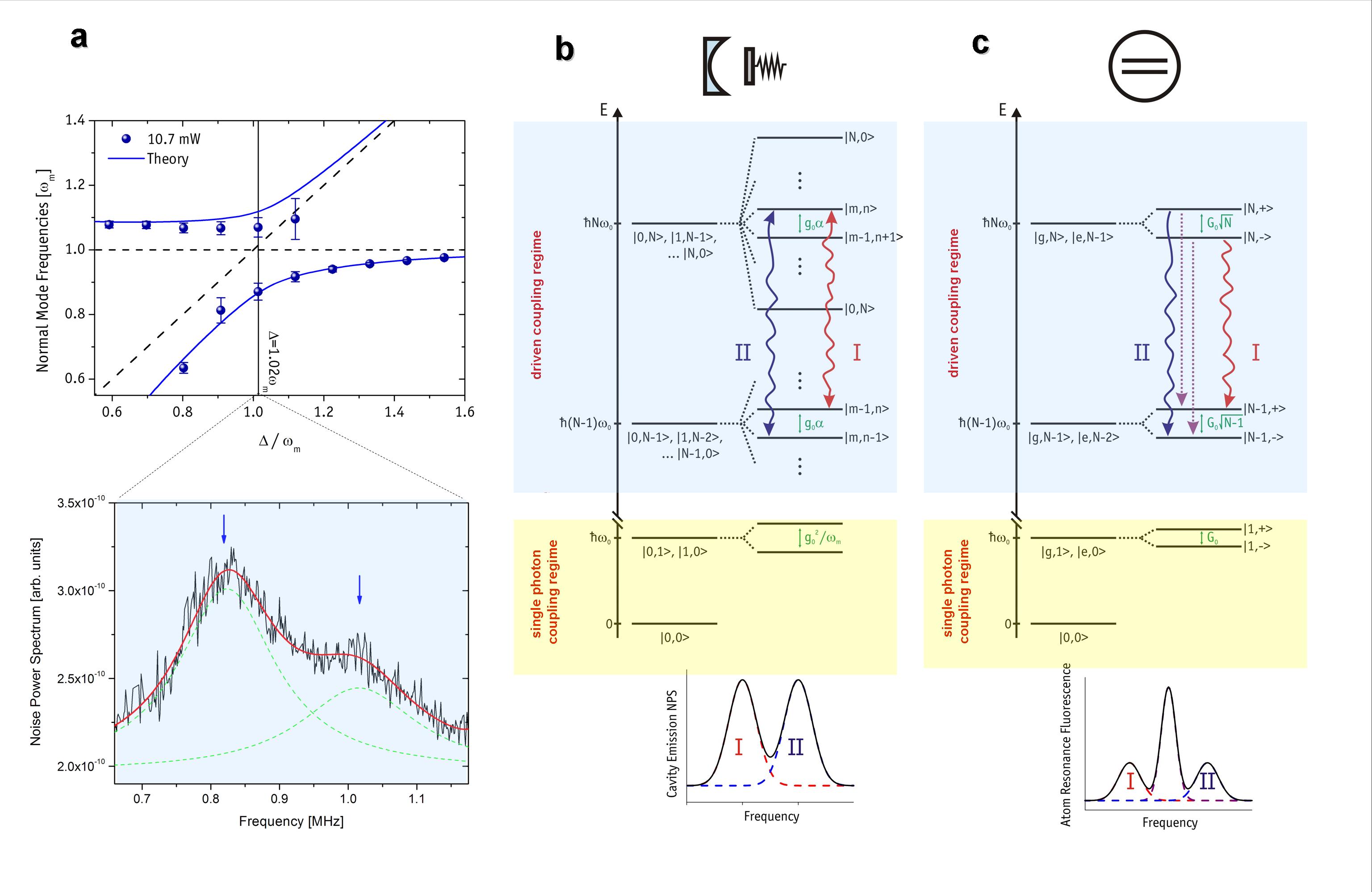}}
\caption{Micromechanics in the strong coupling regime. When the optomechanical coupling rate overcomes the intrinsic decoherence rates, here the cavity decay rate and the mechanical damping rate, the optical mode becomes "dressed" by the mechanical mode. The new dynamics can be described by a set of "normal modes" of a truly hybrid optomechanical system. \textbf{a} Shown are the measured normal mode frequencies of the optomechanical system as a function of cavity detuning (from~[82]). The normal mode frequencies are obtained from fits to the emission spectra of the driven optomechanical cavity, which are measured via sideband homodyne detection of the cooling beam. For far off-resonant driving, the normal modes approach the limiting case of two uncoupled -- one optical and one mechanical -- systems (indicated by the dashed lines). At resonance and for sufficiently strong driving, the normal mode spectrum is split and results in an avoided level crossing. \textbf{b} and \textbf{c} compare a strongly driven optomechanical resonator to a strongly driven two-level system. In the absence of coupling both energy spectra comprise equidistant levels of energy $\hbar N\omega_m$ with degeneracy $N$ and 2, respectively. In the coupled case the degeneracy is lifted. In \textbf{b}, for a cavity in a coherent state with a mean numbers of $\langle n_c\rangle$ photons ($n_c\gg 1$), each level splits up into an $N$-multiplett of dressed states $|m,n\rangle$ that are separated by $g=g_0\sqrt{n_c}$ ($m+n=N$, where $m$, $n$ are the normal mode excitations). Emission of a cavity photon is accompanied only by transitions between dressed states $|m,n\rangle$ and $|m-1,n\rangle$ or $|m,n-1\rangle$. Accordingly, emitted photons have to lie at sideband frequencies $\omega_L+\omega_{\pm}$, where $\omega_{\pm}$ are the frequencies of the normal modes. This gives rise to a doublet structure in the sideband spectrum (bottom) with a splitting. As depicted in \textbf{c}, the same considerations lead to the four different allowed transitions in the case of a strongly coupled two-level system giving rise to the well-known Mollow-triplet in the atom resonance fluorescence. The analogy even holds for strong coupling in the single photon regime (yellow part) where for a two-level system vacuum Rabi splitting is observed—a phenomenon that is predicted for the optomechanical case with a splitting of $g_0^2/\omega_m$.} \label{fig3}
\end{figure*}

\begin{center}
\large\textbf{Strong coupling between optics and mechanics}\\
\end{center}

Based on these latest results there seem to be no fundamental issues that would prevent us from laser cooling mechanical resonators into their quantum ground state, i.e., to initialize the mechanical mode in a pure quantum state, even when starting from room temperature. For sufficiently high mechanical frequencies even brute-force cooling in cryogenic devices is sufficient~\cite{OConnell2010}. In order to complete the quantum optical analogy we have to take into account the fact that the optomechanical system is not perfectly isolated from its environment, particularly, if we compare it to the situation of two optical modes in a typical quantum optics scenario: photons leak out of the cavity at the cavity amplitude decay rate $\kappa$ and the mechanical resonator exchanges phonons with its environment at a rate $\gamma_m$, both leading to decoherence and inhibiting truly joint optomechanical quantum evolution. In order to generate coherent dynamics the time scale on which the optomechanical interaction takes place therefore needs to be smaller than the relevant decoherence time scales in the system. Depending on the specific aim in a given experiment, this requires operation in the large cooperativity regime ($g>\sqrt{\kappa\cdot\gamma_m}$)~\cite{Clerk2008,Hammerer2009} or even in the strong coupling regime ($g>\kappa,\gamma_m$)~\cite{Bose1997,Vitali2007,Marshall2003,Paternostro2007}. We have demonstrated the latter in a recent experiment~\cite{Groeblacher2009b} by increasing the optical pump power close to the cavity instability. The obtained optomechanical coupling strength was sufficiently large to observe normal-mode splitting~\cite{Marquardt2007,Dobrindt2008} as an unambiguous signature of the strong coupling regime (see Fig.\ 3). The relevance of this regime can be seen by an analogy from atomic physics, where atoms can become "dressed" with the interacting photons~\cite{Dalibard1985}. Such dressed states are joint atom-photon states that are inherently entangled and are therefore an important ingredient for quantum state preparation~\cite{Haroche2006}. In our case, the optomechanical dynamics in the strong coupling regime allows for a similar interpretation, because the system's normal modes establish a new set of dynamical variables that cannot be ascribed to either the cavity field or the mechanical resonator but are true hybrid optomechanical degrees of freedom, i.e., optical states dressed by the mechanical resonator (or vice versa). Interestingly, in such an analogy the observed normal-mode splitting can be viewed as spectral splitting from dressed-mode excitations, which are well-known as "Mollow-triplet" in atomic resonance fluorescence~\cite{Groeblacher2009b}. Our experiment demonstrates that the strong coupling regime is accessible with present-day optomechanical devices. However, because it has been performed at a room temperature environment all possible quantum effects are suppressed. A next experiment therefore has to combine strong optomechanical coupling with mechanical devices close to their quantum ground state in order to show the generation and control of true mechanical quantum states.\\

\begin{center}
\large\textbf{Towards Single-Photon Optomechanics}\\
\end{center}

Thus far we have restricted our discussion to quantum optomechanics of a driven optomechanical cavity. On a final note we want to stress that optomechanical effects at the single-photon level~\cite{Akram2010} provide access to the so-far unexplored regime of nonlinear optomechanical interactions. Again, the analogy to quantum atom optics proves helpful: while the energy spectrum of a driven optomechanical system is reminiscent of a two-level atom that is excited by a strong laser field (hence the analogy to the Mollow triplet; see above), the single-photon regime allows state-dependent addressing of individual levels of the optomechanical system (Fig.\ 3) and therefore resembles an optomechanical analogy of vacuum Rabi splitting. As is known, for example, from the field of cavity quantum electrodynamics (cavity-QED), such couplings allow for an even larger variety of protocols for quantum state engineering~\cite{Haroche2006}. The heart of this regime is the generation of strong single-photon nonlinearities in the sense of "strong coupling". Current single-photon coupling rates, $g_0$, in optomechanical systems have not yet reached this regime. For example, most present micro-optomechanical geometries reach values of $g_0$ that are several orders of magnitude smaller than the cavity decay rate $\kappa$. Only recently a few promising candidate systems have been introduced in form of coupled microdisc cavities~\cite{Lin2009} and of photonic waveguide structures~\cite{Li2008,Eichenfield2009,Roels2009}. Here, the radiation pressure forces depend on optical near-field effects that give rise to a significant enhancement of $g_0$. What prevents operation in the single-photon optomechanics regime in these structures thus far is the optical quality, which limits the achievable cavity decay rates. However, the steady progress in the fabrication process may allow us to overcome this limitation soon and give rise to a new realization of cavity-QED in the strong, nonlinear coupling regime, where single atoms are replaced by single mechanical oscillators.\\

\begin{center}
\large\textbf{Summary and Outlook}\\
\end{center}

The recent experiments involving nanomechanics and micromechanics show in an impressive way that the quantum regime of these systems can be exploited with a broad variety of approaches~\cite{Cho2010}. We have provided a brief glance into the field of quantum optomechanics, where quantum optics serves as a particularly helpful toolbox for achieving coherent control over quantum states of mechanical resonators.

While we could discuss only a few examples, many experiments worldwide are underway to exploit these fascinating ideas and their applications. For example, extending the established technology of mechanical sensing to the quantum regime of mechanical resonators may allow new sensing capabilities at or even beyond the quantum limit. Already today, "classical" mechanical resonators reach sensitivities that can resolve masses in the yoctogram ($10^{-24}$~kg) regime~\cite{Jensen2008,Naik2009}, forces in the zeptonewton ($10^{-21}$~N) regime~\cite{Mamin2001}, or displacements in the attometer ($10^{-18}$~m) regime~\cite{Arcizet2006a}, which in turn allowed the measurement of single electron spins~\cite{Rugar2004} of the Casimir force~\cite{Munday2009} or of persistent currents~\cite{Bleszynski-Jayich2009}. It is an intriguing question to ask to what extent one can push these current performances by entering the quantum regime, e.g., in form of squeezed mechanical states~\cite{Mari2009,Jaehne2009}. Another remarkable feature of nano- and micromechanical resonators is their versatility in coupling to many different physical systems, which can be achieved by functionalizing the mechanical object. In combination with their on-chip integrability they become a unique candidate as transducer systems for quantum information processing that enables large-scale coupling between otherwise incompatible quantum systems~\cite{Rabl2009c,Cleland2004}. Several proposals and experiments are attempting to establish an interface between mechanical resonators and atomic ensembles or even single atoms, which would allow to bring in additional concepts from atomic physics~\cite{Hammerer2009b,Hammerer2009,Treutlein2007,Hammerer2010,Wallquist2010}. Finally, preparing quantum superposition states of massive mechanical objects that can contain up to $10^{20}$ atoms opens up a new avenue for novel fundamental tests of macroscopic quantum physics~\cite{Arndt2009,Leggett2002a}, including decoherence at the quantum-classical transition~\cite{Zurek1991}, tests of so-called collapse models~\cite{Marshall2003,Penrose2000,Diosi2000,Adler2009}, or analogues of Schr\"odinger's cat involving living biological systems~\cite{Romero-Isart2010}.

We have only begun to enter the realm of mechanical quantum systems, and these ideas are just examples for their immensely broad spectrum of possible applications. Quantum optomechanics is a promising route to achieve optical control over the mechanical quantum regime. It also provides a toolbox that can eventually be transferred to other implementations of nano- and micromechanics.\\

\begin{acknowledgements}
We acknowledge insightful discussions with our many colleagues all over the world. Our work is supported by the Austrian Science Fund (FWF projects START, FOCUS, P19539, L426), by the $7^{\mathrm{th}}$ Framework Programme of the European Commission (MINOS), by the European Research Council (ERC Starting Grant), and by the Foundational Questions Institute (FQXi). SG is a member of the FWF doctoral program Complex Quantum Systems (W1210) and acknowledges support by the Austrian Academy of Sciences, NK acknowledges support by the Alexander von Humboldt Foundation.
\end{acknowledgements}


\begin{thebibliography}{100}%
\makeatletter
\providecommand \@ifxundefined [1]{%
 \ifx #1\undefined \expandafter \@firstoftwo
 \else \expandafter \@secondoftwo
\fi
}%
\providecommand \@ifnum [1]{%
 \ifnum #1\expandafter \@firstoftwo
 \else \expandafter \@secondoftwo
\fi
}%
\providecommand \enquote [1]{``#1''}%
\providecommand \bibnamefont  [1]{#1}%
\providecommand \bibfnamefont [1]{#1}%
\providecommand \citenamefont [1]{#1}%
\providecommand\href[0]{\@sanitize\@href}%
\providecommand\@href[1]{\endgroup\@@startlink{#1}\endgroup\@@href}%
\providecommand\@@href[1]{#1\@@endlink}%
\providecommand \@sanitize [0]{\begingroup\catcode`\&12\catcode`\#12\relax}%
\@ifxundefined \pdfoutput {\@firstoftwo}{%
 \@ifnum{\z@=\pdfoutput}{\@firstoftwo}{\@secondoftwo}%
}{%
 \providecommand\@@startlink[1]{\leavevmode}%
 \providecommand\@@endlink[0]{}%
}{%
 \providecommand\@@startlink[1]{%
  \leavevmode
  \pdfstartlink
   attr{/Border[0 0 1 ]/H/I/C[0 1 1]}%
   user{/Subtype/Link/A<</Type/Action/S/URI/URI(#1)>>}%
  \relax
 }%
 \providecommand\@@endlink[0]{\pdfendlink}%
}%
\providecommand \url  [0]{\begingroup\@sanitize \@url }%
\providecommand \@url [1]{\endgroup\@href {#1}{\urlprefix}}%
\providecommand \urlprefix [0]{URL }%
\providecommand \Eprint[0]{\href }%
\@ifxundefined \urlstyle {%
  \providecommand \doi [1]{doi:\discretionary{}{}{}#1}%
}{%
  \providecommand \doi [0]{doi:\discretionary{}{}{}\begingroup
  \urlstyle{rm}\Url }%
}%
\providecommand \doibase [0]{http://dx.doi.org/}%
\providecommand \Doi[1]{\href{\doibase#1}}%
\providecommand \bibAnnote [3]{%
  \BibitemShut{#1}%
  \begin{quotation}\noindent
    \textsc{Key:}\ #2\\\textsc{Annotation:}\ #3%
  \end{quotation}%
}%
\providecommand \bibAnnoteFile [2]{%
  \IfFileExists{#2}{\bibAnnote {#1} {#2} {\input{#2}}}{}%
}%
\providecommand \typeout [0]{\immediate \write \m@ne }%
\providecommand \selectlanguage [0]{\@gobble}%
\providecommand \bibinfo [0]{\@secondoftwo}%
\providecommand \bibfield [0]{\@secondoftwo}%
\providecommand \translation [1]{[#1]}%
\providecommand \BibitemOpen[0]{}%
\providecommand \bibitemStop [0]{}%
\providecommand \bibitemNoStop [0]{.\EOS\space}%
\providecommand \EOS [0]{\spacefactor3000\relax}%
\providecommand \BibitemShut [1]{\csname bibitem#1\endcsname}%
\bibitem{Southwell2008}%
  \BibitemOpen
  \bibfield{author}{%
  \bibinfo {author} {\bibfnamefont{K.}~\bibnamefont{Southwell}},\ }%
  \bibfield{journal}{%
  \bibinfo {journal} {Nature}\ }%
  \textbf{\bibinfo {volume} {453}},\ \bibinfo {pages} {1003} (\bibinfo {year}
  {2008})%
  \bibAnnoteFile{NoStop}{Southwell2008}%
\bibitem{Osborne2008}%
  \BibitemOpen
  \bibfield{author}{%
  \bibinfo {author} {\bibfnamefont{I.}~\bibnamefont{Osborne}}\ and\ \bibinfo
  {author} {\bibfnamefont{R.}~\bibnamefont{Coontz}},\ }%
  \bibfield{journal}{%
  \bibinfo {journal} {Science}\ }%
  \textbf{\bibinfo {volume} {319}},\ \bibinfo {pages} {1201} (\bibinfo {year}
  {2008})%
  \bibAnnoteFile{NoStop}{Osborne2008}%
\bibitem{Zoller2005}%
  \BibitemOpen
  \bibfield{author}{%
  \bibinfo {author} {\bibfnamefont{P.}~\bibnamefont{Zoller}}, \bibinfo {author}
  {\bibfnamefont{T.}~\bibnamefont{Beth}}, \bibinfo {author}
  {\bibfnamefont{D.}~\bibnamefont{Binosi}}, \bibinfo {author}
  {\bibfnamefont{R.}~\bibnamefont{Blatt}}, \bibinfo {author}
  {\bibfnamefont{H.}~\bibnamefont{Briegel}}, \bibinfo {author}
  {\bibfnamefont{D.}~\bibnamefont{Bruss}}, \bibinfo {author}
  {\bibfnamefont{T.}~\bibnamefont{Calarco}}, \bibinfo {author}
  {\bibfnamefont{J.}~\bibnamefont{Cirac}}, \bibinfo {author}
  {\bibfnamefont{D.}~\bibnamefont{Deutsch}}, \bibinfo {author}
  {\bibfnamefont{J.}~\bibnamefont{Eisert}}, \bibinfo {author}
  {\bibfnamefont{A.}~\bibnamefont{Ekert}}, \bibinfo {author}
  {\bibfnamefont{C.}~\bibnamefont{Fabre}}, \bibinfo {author}
  {\bibfnamefont{N.}~\bibnamefont{Gisin}}, \bibinfo {author}
  {\bibfnamefont{P.}~\bibnamefont{Grangiere}}, \bibinfo {author}
  {\bibfnamefont{M.}~\bibnamefont{Grassl}}, \bibinfo {author}
  {\bibfnamefont{S.}~\bibnamefont{Haroche}}, \bibinfo {author}
  {\bibfnamefont{A.}~\bibnamefont{Imamoglu}}, \bibinfo {author}
  {\bibfnamefont{A.}~\bibnamefont{Karlson}}, \bibinfo {author}
  {\bibfnamefont{J.}~\bibnamefont{Kempe}}, \bibinfo {author}
  {\bibfnamefont{L.}~\bibnamefont{Kouwenhoven}}, \bibinfo {author}
  {\bibfnamefont{S.}~\bibnamefont{Kr\"oll}}, \bibinfo {author}
  {\bibfnamefont{G.}~\bibnamefont{Leuchs}}, \bibinfo {author}
  {\bibfnamefont{M.}~\bibnamefont{Lewenstein}}, \bibinfo {author}
  {\bibfnamefont{D.}~\bibnamefont{Loss}}, \bibinfo {author}
  {\bibfnamefont{N.}~\bibnamefont{L\"utkenhaus}}, \bibinfo {author}
  {\bibfnamefont{S.}~\bibnamefont{Massar}}, \bibinfo {author}
  {\bibfnamefont{J.}~\bibnamefont{Mooij}}, \bibinfo {author}
  {\bibfnamefont{M.}~\bibnamefont{Plenio}}, \bibinfo {author}
  {\bibfnamefont{E.}~\bibnamefont{Polzik}}, \bibinfo {author}
  {\bibfnamefont{S.}~\bibnamefont{Popescu}}, \bibinfo {author}
  {\bibfnamefont{G.}~\bibnamefont{Rempe}}, \bibinfo {author}
  {\bibfnamefont{A.}~\bibnamefont{Sergienko}}, \bibinfo {author}
  {\bibfnamefont{D.}~\bibnamefont{Suter}}, \bibinfo {author}
  {\bibfnamefont{J.}~\bibnamefont{Twamley}}, \bibinfo {author}
  {\bibfnamefont{G.}~\bibnamefont{Wendin}}, \bibinfo {author}
  {\bibfnamefont{R.}~\bibnamefont{Werner}}, \bibinfo {author}
  {\bibfnamefont{A.}~\bibnamefont{Winter}}, \bibinfo {author}
  {\bibfnamefont{J.}~\bibnamefont{Wrachtrup}},\ and\ \bibinfo {author}
  {\bibfnamefont{A.}~\bibnamefont{Zeilinger}},\ }%
  \bibfield{journal}{%
  \bibinfo {journal} {Eur.\ Phys.\ J.\ D}\ }%
  \textbf{\bibinfo {volume} {36}},\ \bibinfo {pages} {203} (\bibinfo {year}
  {2005})%
  \bibAnnoteFile{NoStop}{Zoller2005}%
\bibitem{Aspelmeyer2008b}%
  \BibitemOpen
  \bibfield{author}{%
  \bibinfo {author} {\bibfnamefont{M.}~\bibnamefont{Aspelmeyer}}\ and\ \bibinfo
  {author} {\bibfnamefont{A.}~\bibnamefont{Zeilinger}},\ }%
  \bibfield{journal}{%
  \bibinfo {journal} {Physics World}\ }%
  \textbf{\bibinfo {volume} {21}} (\bibinfo {year} {2008})%
  \bibAnnoteFile{NoStop}{Aspelmeyer2008b}%
\bibitem{Blatt2008}%
  \BibitemOpen
  \bibfield{author}{%
  \bibinfo {author} {\bibfnamefont{R.}~\bibnamefont{Blatt}}\ and\ \bibinfo
  {author} {\bibfnamefont{D.}~\bibnamefont{Wineland}},\ }%
  \bibfield{journal}{%
  \bibinfo {journal} {Nature}\ }%
  \textbf{\bibinfo {volume} {453}},\ \bibinfo {pages} {1008} (\bibinfo {year}
  {2008})%
  \bibAnnoteFile{NoStop}{Blatt2008}%
\bibitem{Jost2009}%
  \BibitemOpen
  \bibfield{author}{%
  \bibinfo {author} {\bibfnamefont{J.~D.}\ \bibnamefont{Jost}}, \bibinfo
  {author} {\bibfnamefont{J.~P.}\ \bibnamefont{Home}}, \bibinfo {author}
  {\bibfnamefont{J.~M.}\ \bibnamefont{Amini}}, \bibinfo {author}
  {\bibfnamefont{D.}~\bibnamefont{Hanneke}}, \bibinfo {author}
  {\bibfnamefont{R.}~\bibnamefont{Ozeri}}, \bibinfo {author}
  {\bibfnamefont{C.}~\bibnamefont{Langer}}, \bibinfo {author}
  {\bibfnamefont{J.~J.}\ \bibnamefont{Bollinger}}, \bibinfo {author}
  {\bibfnamefont{D.}~\bibnamefont{Leibfried}},\ and\ \bibinfo {author}
  {\bibfnamefont{D.~J.}\ \bibnamefont{Wineland}},\ }%
  \bibfield{journal}{%
  \bibinfo {journal} {Nature}\ }%
  \textbf{\bibinfo {volume} {459}},\ \bibinfo {pages} {683} (\bibinfo {year}
  {2009})%
  \bibAnnoteFile{NoStop}{Jost2009}%
\bibitem{Cleland1996}%
  \BibitemOpen
  \bibfield{author}{%
  \bibinfo {author} {\bibfnamefont{A.~N.}\ \bibnamefont{Cleland}}\ and\
  \bibinfo {author} {\bibfnamefont{M.~L.}\ \bibnamefont{Roukes}},\ }%
  \bibfield{journal}{%
  \bibinfo {journal} {Appl.\ Phys.\ Lett.}\ }%
  \textbf{\bibinfo {volume} {69}},\ \bibinfo {pages} {2653} (\bibinfo {year}
  {1996})%
  \bibAnnoteFile{NoStop}{Cleland1996}%
\bibitem{Cleland1998}%
  \BibitemOpen
  \bibfield{author}{%
  \bibinfo {author} {\bibfnamefont{A.~N.}\ \bibnamefont{Cleland}}\ and\
  \bibinfo {author} {\bibfnamefont{M.~L.}\ \bibnamefont{Roukes}},\ }%
  \bibfield{journal}{%
  \bibinfo {journal} {Nature}\ }%
  \textbf{\bibinfo {volume} {392}},\ \bibinfo {pages} {160} (\bibinfo {year}
  {1998})%
  \bibAnnoteFile{NoStop}{Cleland1998}%
\bibitem{Schwab2005}%
  \BibitemOpen
  \bibfield{author}{%
  \bibinfo {author} {\bibfnamefont{K.~C.}\ \bibnamefont{Schwab}}\ and\ \bibinfo
  {author} {\bibfnamefont{M.~L.}\ \bibnamefont{Roukes}},\ }%
  \bibfield{journal}{%
  \bibinfo {journal} {Physics Today}\ }%
  \textbf{\bibinfo {volume} {58}},\ \bibinfo {pages} {36} (\bibinfo {month}
  {July}\ \bibinfo {year} {2005})%
  \bibAnnoteFile{NoStop}{Schwab2005}%
\bibitem{Cho2010}%
  \BibitemOpen
  \bibfield{author}{%
  \bibinfo {author} {\bibfnamefont{A.}~\bibnamefont{Cho}},\ }%
  \bibfield{journal}{%
  \bibinfo {journal} {Science}\ }%
  \textbf{\bibinfo {volume} {327}},\ \bibinfo {pages} {516} (\bibinfo {year}
  {2010})%
  \bibAnnoteFile{NoStop}{Cho2010}%
\bibitem{Rugar2004}%
  \BibitemOpen
  \bibfield{author}{%
  \bibinfo {author} {\bibfnamefont{D.}~\bibnamefont{Rugar}}, \bibinfo {author}
  {\bibfnamefont{R.}~\bibnamefont{Budakian}}, \bibinfo {author}
  {\bibfnamefont{H.~J.}\ \bibnamefont{Mamin}},\ and\ \bibinfo {author}
  {\bibfnamefont{B.~W.}\ \bibnamefont{Chui}},\ }%
  \bibfield{journal}{%
  \bibinfo {journal} {Nature}\ }%
  \textbf{\bibinfo {volume} {430}},\ \bibinfo {pages} {329} (\bibinfo {year}
  {2004})%
  \bibAnnoteFile{NoStop}{Rugar2004}%
\bibitem{LaHaye2004}%
  \BibitemOpen
  \bibfield{author}{%
  \bibinfo {author} {\bibfnamefont{M.~D.}\ \bibnamefont{LaHaye}}, \bibinfo
  {author} {\bibfnamefont{O.}~\bibnamefont{Buu}}, \bibinfo {author}
  {\bibfnamefont{B.}~\bibnamefont{Camarota}},\ and\ \bibinfo {author}
  {\bibfnamefont{K.~C.}\ \bibnamefont{Schwab}},\ }%
  \bibfield{journal}{%
  \bibinfo {journal} {Science}\ }%
  \textbf{\bibinfo {volume} {304}},\ \bibinfo {pages} {74} (\bibinfo {year}
  {2004})%
  \bibAnnoteFile{NoStop}{LaHaye2004}%
\bibitem{LaHaye2009}%
  \BibitemOpen
  \bibfield{author}{%
  \bibinfo {author} {\bibfnamefont{M.~D.}\ \bibnamefont{LaHaye}}, \bibinfo
  {author} {\bibfnamefont{J.}~\bibnamefont{Suh}}, \bibinfo {author}
  {\bibfnamefont{P.~M.}\ \bibnamefont{Echternach}}, \bibinfo {author}
  {\bibfnamefont{K.~C.}\ \bibnamefont{Schwab}},\ and\ \bibinfo {author}
  {\bibfnamefont{M.~L.}\ \bibnamefont{Roukes}},\ }%
  \bibfield{journal}{%
  \bibinfo {journal} {Nature}\ }%
  \textbf{\bibinfo {volume} {459}},\ \bibinfo {pages} {960} (\bibinfo {year}
  {2009})%
  \bibAnnoteFile{NoStop}{LaHaye2009}%
\bibitem{OConnell2010}%
  \BibitemOpen
  \bibfield{author}{%
  \bibinfo {author} {\bibfnamefont{A.~D.}\ \bibnamefont{O'Connell}}, \bibinfo
  {author} {\bibfnamefont{M.}~\bibnamefont{Hofheinz}}, \bibinfo {author}
  {\bibfnamefont{M.}~\bibnamefont{Ansmann}}, \bibinfo {author}
  {\bibfnamefont{R.~C.}\ \bibnamefont{Bialczak}}, \bibinfo {author}
  {\bibfnamefont{M.}~\bibnamefont{Lenander}}, \bibinfo {author}
  {\bibfnamefont{E.}~\bibnamefont{Lucero}}, \bibinfo {author}
  {\bibfnamefont{M.}~\bibnamefont{Neeley}}, \bibinfo {author}
  {\bibfnamefont{D.}~\bibnamefont{Sank}}, \bibinfo {author}
  {\bibfnamefont{H.}~\bibnamefont{Wang}}, \bibinfo {author}
  {\bibfnamefont{M.}~\bibnamefont{Weides}}, \bibinfo {author}
  {\bibfnamefont{J.}~\bibnamefont{Wenner}}, \bibinfo {author}
  {\bibfnamefont{J.~M.}\ \bibnamefont{Martinis}},\ and\ \bibinfo {author}
  {\bibfnamefont{A.~N.}\ \bibnamefont{Cleland}},\ }%
  \bibfield{journal}{%
  \bibinfo {journal} {Nature}\ }%
  \textbf{\bibinfo {volume} {464}},\ \bibinfo {pages} {697} (\bibinfo {year}
  {2010})%
  \bibAnnoteFile{NoStop}{OConnell2010}%
\bibitem{Kippenberg2005}%
  \BibitemOpen
  \bibfield{author}{%
  \bibinfo {author} {\bibfnamefont{T.}~\bibnamefont{Kippenberg}}, \bibinfo
  {author} {\bibfnamefont{H.}~\bibnamefont{Rokhsari}}, \bibinfo {author}
  {\bibfnamefont{T.}~\bibnamefont{Carmon}}, \bibinfo {author}
  {\bibfnamefont{A.}~\bibnamefont{Scherer}},\ and\ \bibinfo {author}
  {\bibfnamefont{K.}~\bibnamefont{Vahala}},\ }%
  \bibfield{journal}{%
  \bibinfo {journal} {Phys.\ Rev.\ Lett.}\ }%
  \textbf{\bibinfo {volume} {95}},\ \bibinfo {pages} {033901} (\bibinfo {month}
  {07}\ \bibinfo {year} {2005})%
  \bibAnnoteFile{NoStop}{Kippenberg2005}%
\bibitem{Gigan2006}%
  \BibitemOpen
  \bibfield{author}{%
  \bibinfo {author} {\bibfnamefont{S.}~\bibnamefont{Gigan}}, \bibinfo {author}
  {\bibfnamefont{H.~R.}\ \bibnamefont{B\"ohm}}, \bibinfo {author}
  {\bibfnamefont{M.}~\bibnamefont{Paternostro}}, \bibinfo {author}
  {\bibfnamefont{F.}~\bibnamefont{Blaser}}, \bibinfo {author}
  {\bibfnamefont{G.}~\bibnamefont{Langer}}, \bibinfo {author}
  {\bibfnamefont{J.~B.}\ \bibnamefont{Hertzberg}}, \bibinfo {author}
  {\bibfnamefont{K.~C.}\ \bibnamefont{Schwab}}, \bibinfo {author}
  {\bibfnamefont{D.}~\bibnamefont{B\"auerle}}, \bibinfo {author}
  {\bibfnamefont{M.}~\bibnamefont{Aspelmeyer}},\ and\ \bibinfo {author}
  {\bibfnamefont{A.}~\bibnamefont{Zeilinger}},\ }%
  \bibfield{journal}{%
  \bibinfo {journal} {Nature}\ }%
  \textbf{\bibinfo {volume} {444}},\ \bibinfo {pages} {67} (\bibinfo {year}
  {2006})%
  \bibAnnoteFile{NoStop}{Gigan2006}%
\bibitem{Arcizet2006b}%
  \BibitemOpen
  \bibfield{author}{%
  \bibinfo {author} {\bibfnamefont{O.}~\bibnamefont{Arcizet}}, \bibinfo
  {author} {\bibfnamefont{P.-F.}\ \bibnamefont{Cohadon}}, \bibinfo {author}
  {\bibfnamefont{T.}~\bibnamefont{Briant}}, \bibinfo {author}
  {\bibfnamefont{M.}~\bibnamefont{Pinard}},\ and\ \bibinfo {author}
  {\bibfnamefont{A.}~\bibnamefont{Heidmann}},\ }%
  \bibfield{journal}{%
  \bibinfo {journal} {Nature}\ }%
  \textbf{\bibinfo {volume} {444}},\ \bibinfo {pages} {71} (\bibinfo {year}
  {2006})%
  \bibAnnoteFile{NoStop}{Arcizet2006b}%
\bibitem{Thompson2008}%
  \BibitemOpen
  \bibfield{author}{%
  \bibinfo {author} {\bibfnamefont{J.~D.}\ \bibnamefont{Thompson}}, \bibinfo
  {author} {\bibfnamefont{B.~M.}\ \bibnamefont{Zwickl}}, \bibinfo {author}
  {\bibfnamefont{A.~M.}\ \bibnamefont{Jayich}}, \bibinfo {author}
  {\bibfnamefont{F.}~\bibnamefont{Marquardt}}, \bibinfo {author}
  {\bibfnamefont{S.~M.}\ \bibnamefont{Girvin}},\ and\ \bibinfo {author}
  {\bibfnamefont{J.~G.~E.}\ \bibnamefont{Harris}},\ }%
  \bibfield{journal}{%
  \bibinfo {journal} {Nature}\ }%
  \textbf{\bibinfo {volume} {452}},\ \bibinfo {pages} {72} (\bibinfo {year}
  {2008})%
  \bibAnnoteFile{NoStop}{Thompson2008}%
\bibitem{Regal2008}%
  \BibitemOpen
  \bibfield{author}{%
  \bibinfo {author} {\bibfnamefont{C.~A.}\ \bibnamefont{Regal}}, \bibinfo
  {author} {\bibfnamefont{J.~D.}\ \bibnamefont{Teufel}},\ and\ \bibinfo
  {author} {\bibfnamefont{K.~W.}\ \bibnamefont{Lehnert}},\ }%
  \bibfield{journal}{%
  \bibinfo {journal} {Nature Phys.}\ }%
  \textbf{\bibinfo {volume} {4}},\ \bibinfo {pages} {555} (\bibinfo {year}
  {2008})%
  \bibAnnoteFile{NoStop}{Regal2008}%
\bibitem{Law1994}%
  \BibitemOpen
  \bibfield{author}{%
  \bibinfo {author} {\bibfnamefont{C.~K.}\ \bibnamefont{Law}},\ }%
  \bibfield{journal}{%
  \bibinfo {journal} {Phys.\ Rev.\ A}\ }%
  \textbf{\bibinfo {volume} {49}},\ \bibinfo {pages} {433} (\bibinfo {month}
  {1}\ \bibinfo {year} {1994})%
  \bibAnnoteFile{NoStop}{Law1994}%
\bibitem{Li2008}%
  \BibitemOpen
  \bibfield{author}{%
  \bibinfo {author} {\bibfnamefont{M.}~\bibnamefont{Li}}, \bibinfo {author}
  {\bibfnamefont{W.~H.~P.}\ \bibnamefont{Pernice}}, \bibinfo {author}
  {\bibfnamefont{C.}~\bibnamefont{Xiong}}, \bibinfo {author}
  {\bibfnamefont{T.}~\bibnamefont{Baehr-Jones}}, \bibinfo {author}
  {\bibfnamefont{M.}~\bibnamefont{Hochberg}},\ and\ \bibinfo {author}
  {\bibfnamefont{H.~X.}\ \bibnamefont{Tang}},\ }%
  \bibfield{journal}{%
  \bibinfo {journal} {Nature}\ }%
  \textbf{\bibinfo {volume} {456}},\ \bibinfo {pages} {480} (\bibinfo {year}
  {2008})%
  \bibAnnoteFile{NoStop}{Li2008}%
\bibitem{Eichenfield2009}%
  \BibitemOpen
  \bibfield{author}{%
  \bibinfo {author} {\bibfnamefont{M.}~\bibnamefont{Eichenfield}}, \bibinfo
  {author} {\bibfnamefont{R.}~\bibnamefont{Camacho}}, \bibinfo {author}
  {\bibfnamefont{J.}~\bibnamefont{Chan}}, \bibinfo {author}
  {\bibfnamefont{K.~J.}\ \bibnamefont{Vahala}},\ and\ \bibinfo {author}
  {\bibfnamefont{O.}~\bibnamefont{Painter}},\ }%
  \bibfield{journal}{%
  \bibinfo {journal} {Nature}\ }%
  \textbf{\bibinfo {volume} {459}},\ \bibinfo {pages} {550} (\bibinfo {year}
  {2009})%
  \bibAnnoteFile{NoStop}{Eichenfield2009}%
\bibitem{Roels2009}%
  \BibitemOpen
  \bibfield{author}{%
  \bibinfo {author} {\bibfnamefont{J.}~\bibnamefont{Roels}}, \bibinfo {author}
  {\bibfnamefont{I.}~\bibnamefont{{De Vlaminck}}}, \bibinfo {author}
  {\bibfnamefont{L.}~\bibnamefont{Lagae}}, \bibinfo {author}
  {\bibfnamefont{B.}~\bibnamefont{Maes}}, \bibinfo {author}
  {\bibfnamefont{D.}~\bibnamefont{{Van Thourhout}}},\ and\ \bibinfo {author}
  {\bibfnamefont{R.}~\bibnamefont{Baets}},\ }%
  \bibfield{journal}{%
  \bibinfo {journal} {Nature Nanotech.}\ }%
  \textbf{\bibinfo {volume} {4}},\ \bibinfo {pages} {510} (\bibinfo {year}
  {2009})%
  \bibAnnoteFile{NoStop}{Roels2009}%
\bibitem{Anetsberger2009}%
  \BibitemOpen
  \bibfield{author}{%
  \bibinfo {author} {\bibfnamefont{G.}~\bibnamefont{Anetsberger}}, \bibinfo
  {author} {\bibfnamefont{O.}~\bibnamefont{Arcizet}}, \bibinfo {author}
  {\bibfnamefont{Q.~P.}\ \bibnamefont{Unterreithmeier}}, \bibinfo {author}
  {\bibfnamefont{R.}~\bibnamefont{Rivi\`{e}re}}, \bibinfo {author}
  {\bibfnamefont{A.}~\bibnamefont{Schliesser}}, \bibinfo {author}
  {\bibfnamefont{E.~M.}\ \bibnamefont{Weig}}, \bibinfo {author}
  {\bibfnamefont{J.~P.}\ \bibnamefont{Kotthaus}},\ and\ \bibinfo {author}
  {\bibfnamefont{T.~J.}\ \bibnamefont{Kippenberg}},\ }%
  \bibfield{journal}{%
  \bibinfo {journal} {Nature Phys.}\ }%
  \textbf{\bibinfo {volume} {5}},\ \bibinfo {pages} {909} (\bibinfo {year}
  {2009})%
  \bibAnnoteFile{NoStop}{Anetsberger2009}%
\bibitem{Braginsky1970}%
  \BibitemOpen
  \bibfield{author}{%
  \bibinfo {author} {\bibfnamefont{V.~B.}\ \bibnamefont{Braginsky}}, \bibinfo
  {author} {\bibfnamefont{A.~B.}\ \bibnamefont{Manukin}},\ and\ \bibinfo
  {author} {\bibfnamefont{M.~Y.}\ \bibnamefont{Tikhonov}},\ }%
  \bibfield{journal}{%
  \bibinfo {journal} {Sov.\ Phys.\ JETP}\ }%
  \textbf{\bibinfo {volume} {31}},\ \bibinfo {pages} {829} (\bibinfo {year}
  {1970})%
  \bibAnnoteFile{NoStop}{Braginsky1970}%
\bibitem{Fabre1994}%
  \BibitemOpen
  \bibfield{author}{%
  \bibinfo {author} {\bibfnamefont{C.}~\bibnamefont{Fabre}}, \bibinfo {author}
  {\bibfnamefont{M.}~\bibnamefont{Pinard}}, \bibinfo {author}
  {\bibfnamefont{S.}~\bibnamefont{Bourzeix}}, \bibinfo {author}
  {\bibfnamefont{A.}~\bibnamefont{Heidmann}}, \bibinfo {author}
  {\bibfnamefont{E.}~\bibnamefont{Giacobino}},\ and\ \bibinfo {author}
  {\bibfnamefont{S.}~\bibnamefont{Reynaud}},\ }%
  \bibfield{journal}{%
  \bibinfo {journal} {Phys.\ Rev.\ A}\ }%
  \textbf{\bibinfo {volume} {49}},\ \bibinfo {pages} {1337} (\bibinfo {year}
  {1994})%
  \bibAnnoteFile{NoStop}{Fabre1994}%
\bibitem{Mancini1994}%
  \BibitemOpen
  \bibfield{author}{%
  \bibinfo {author} {\bibfnamefont{S.}~\bibnamefont{Mancini}}\ and\ \bibinfo
  {author} {\bibfnamefont{P.}~\bibnamefont{Tombesi}},\ }%
  \bibfield{journal}{%
  \bibinfo {journal} {Phys.\ Rev.\ A}\ }%
  \textbf{\bibinfo {volume} {49}},\ \bibinfo {pages} {4055} (\bibinfo {month}
  {5}\ \bibinfo {year} {1994})%
  \bibAnnoteFile{NoStop}{Mancini1994}%
\bibitem{Milburn1994}%
  \BibitemOpen
  \bibfield{author}{%
  \bibinfo {author} {\bibfnamefont{G.~J.}\ \bibnamefont{Milburn}}, \bibinfo
  {author} {\bibfnamefont{K.}~\bibnamefont{Jacobs}},\ and\ \bibinfo {author}
  {\bibfnamefont{D.~F.}\ \bibnamefont{Walls}},\ }%
  \bibfield{journal}{%
  \bibinfo {journal} {Phys.\ Rev.\ A}\ }%
  \textbf{\bibinfo {volume} {50}},\ \bibinfo {pages} {5256} (\bibinfo {month}
  {Dec}\ \bibinfo {year} {1994})%
  \bibAnnoteFile{NoStop}{Milburn1994}%
\bibitem{Pinard1995}%
  \BibitemOpen
  \bibfield{author}{%
  \bibinfo {author} {\bibfnamefont{M.}~\bibnamefont{Pinard}}, \bibinfo {author}
  {\bibfnamefont{C.}~\bibnamefont{Fabre}},\ and\ \bibinfo {author}
  {\bibfnamefont{A.}~\bibnamefont{Heidmann}},\ }%
  \bibfield{journal}{%
  \bibinfo {journal} {Phys.\ Rev.\ A}\ }%
  \textbf{\bibinfo {volume} {51}},\ \bibinfo {pages} {2443} (\bibinfo {year}
  {1995})%
  \bibAnnoteFile{NoStop}{Pinard1995}%
\bibitem{Mancini1998}%
  \BibitemOpen
  \bibfield{author}{%
  \bibinfo {author} {\bibfnamefont{S.}~\bibnamefont{Mancini}}, \bibinfo
  {author} {\bibfnamefont{D.}~\bibnamefont{Vitali}},\ and\ \bibinfo {author}
  {\bibfnamefont{P.}~\bibnamefont{Tombesi}},\ }%
  \bibfield{journal}{%
  \bibinfo {journal} {Phys.\ Rev.\ Lett.}\ }%
  \textbf{\bibinfo {volume} {80}},\ \bibinfo {pages} {688} (\bibinfo {year}
  {1998})%
  \bibAnnoteFile{NoStop}{Mancini1998}%
\bibitem{Cohadon1999}%
  \BibitemOpen
  \bibfield{author}{%
  \bibinfo {author} {\bibfnamefont{P.-F.}\ \bibnamefont{Cohadon}}, \bibinfo
  {author} {\bibfnamefont{A.}~\bibnamefont{Heidmann}},\ and\ \bibinfo {author}
  {\bibfnamefont{M.}~\bibnamefont{Pinard}},\ }%
  \bibfield{journal}{%
  \bibinfo {journal} {Phys.\ Rev.\ Lett.}\ }%
  \textbf{\bibinfo {volume} {83}},\ \bibinfo {pages} {3174} (\bibinfo {month}
  {10}\ \bibinfo {year} {1999})%
  \bibAnnoteFile{NoStop}{Cohadon1999}%
\bibitem{Bose1997}%
  \BibitemOpen
  \bibfield{author}{%
  \bibinfo {author} {\bibfnamefont{S.}~\bibnamefont{Bose}}, \bibinfo {author}
  {\bibfnamefont{K.}~\bibnamefont{Jacobs}},\ and\ \bibinfo {author}
  {\bibfnamefont{P.~L.}\ \bibnamefont{Knight}},\ }%
  \bibfield{journal}{%
  \bibinfo {journal} {Phys.\ Rev.\ A}\ }%
  \textbf{\bibinfo {volume} {56}},\ \bibinfo {pages} {4175} (\bibinfo {year}
  {1997})%
  \bibAnnoteFile{NoStop}{Bose1997}%
\bibitem{Mancini1997}%
  \BibitemOpen
  \bibfield{author}{%
  \bibinfo {author} {\bibfnamefont{S.}~\bibnamefont{Mancini}}, \bibinfo
  {author} {\bibfnamefont{V.~I.}\ \bibnamefont{Man'ko}},\ and\ \bibinfo
  {author} {\bibfnamefont{P.}~\bibnamefont{Tombesi}},\ }%
  \bibfield{journal}{%
  \bibinfo {journal} {Phys.\ Rev.\ A}\ }%
  \textbf{\bibinfo {volume} {55}},\ \bibinfo {pages} {3042} (\bibinfo {year}
  {1997})%
  \bibAnnoteFile{NoStop}{Mancini1997}%
\bibitem{Dorsel1983}%
  \BibitemOpen
  \bibfield{author}{%
  \bibinfo {author} {\bibfnamefont{A.}~\bibnamefont{Dorsel}}, \bibinfo {author}
  {\bibfnamefont{J.}~\bibnamefont{McCullen}}, \bibinfo {author}
  {\bibfnamefont{P.}~\bibnamefont{Meystre}}, \bibinfo {author}
  {\bibfnamefont{E.}~\bibnamefont{Vignes}},\ and\ \bibinfo {author}
  {\bibfnamefont{H.}~\bibnamefont{Walther}},\ }%
  \bibfield{journal}{%
  \bibinfo {journal} {Phys.\ Rev.\ Lett.}\ }%
  \textbf{\bibinfo {volume} {51}},\ \bibinfo {pages} {1550} (\bibinfo {month}
  {10}\ \bibinfo {year} {1983})%
  \bibAnnoteFile{NoStop}{Dorsel1983}%
\bibitem{Metzger2004}%
  \BibitemOpen
  \bibfield{author}{%
  \bibinfo {author} {\bibfnamefont{C.~H.}\ \bibnamefont{Metzger}}\ and\
  \bibinfo {author} {\bibfnamefont{K.}~\bibnamefont{Karrai}},\ }%
  \bibfield{journal}{%
  \bibinfo {journal} {Nature}\ }%
  \textbf{\bibinfo {volume} {432}},\ \bibinfo {pages} {1002} (\bibinfo {month}
  {12}\ \bibinfo {year} {2004})%
  \bibAnnoteFile{NoStop}{Metzger2004}%
\bibitem{Kippenberg2008}%
  \BibitemOpen
  \bibfield{author}{%
  \bibinfo {author} {\bibfnamefont{T.~J.}\ \bibnamefont{Kippenberg}}\ and\
  \bibinfo {author} {\bibfnamefont{K.~J.}\ \bibnamefont{Vahala}},\ }%
  \bibfield{journal}{%
  \bibinfo {journal} {Science}\ }%
  \textbf{\bibinfo {volume} {321}},\ \bibinfo {pages} {1172} (\bibinfo {year}
  {2008})%
  \bibAnnoteFile{NoStop}{Kippenberg2008}%
\bibitem{Aspelmeyer2008}%
  \BibitemOpen
  \bibfield{author}{%
  \bibinfo {author} {\bibfnamefont{M.}~\bibnamefont{Aspelmeyer}}\ and\ \bibinfo
  {author} {\bibfnamefont{K.~C.}\ \bibnamefont{Schwab}},\ }%
  \bibfield{journal}{%
  \bibinfo {journal} {New J.\ Phys.}\ }%
  \textbf{\bibinfo {volume} {10}},\ \bibinfo {pages} {095001} (\bibinfo {year}
  {2008})%
  \bibAnnoteFile{NoStop}{Aspelmeyer2008}%
\bibitem{Favero2009}%
  \BibitemOpen
  \bibfield{author}{%
  \bibinfo {author} {\bibfnamefont{I.}~\bibnamefont{Favero}}\ and\ \bibinfo
  {author} {\bibfnamefont{K.}~\bibnamefont{Karrai}},\ }%
  \bibfield{journal}{%
  \bibinfo {journal} {Nature Photon.}\ }%
  \textbf{\bibinfo {volume} {3}},\ \bibinfo {pages} {201} (\bibinfo {year}
  {2009})%
  \bibAnnoteFile{NoStop}{Favero2009}%
\bibitem{Marquardt2009}%
  \BibitemOpen
  \bibfield{author}{%
  \bibinfo {author} {\bibfnamefont{F.}~\bibnamefont{Marquardt}}\ and\ \bibinfo
  {author} {\bibfnamefont{S.~M.}\ \bibnamefont{Girvin}},\ }%
  \bibfield{journal}{%
  \bibinfo {journal} {Physics}\ }%
  \textbf{\bibinfo {volume} {2}},\ \bibinfo {pages} {40} (\bibinfo {year}
  {2009})%
  \bibAnnoteFile{NoStop}{Marquardt2009}%
\bibitem{Genes2009}%
  \BibitemOpen
  \bibfield{author}{%
  \bibinfo {author} {\bibfnamefont{C.}~\bibnamefont{Genes}}, \bibinfo {author}
  {\bibfnamefont{A.}~\bibnamefont{Mari}}, \bibinfo {author}
  {\bibfnamefont{D.}~\bibnamefont{Vitali}},\ and\ \bibinfo {author}
  {\bibfnamefont{P.}~\bibnamefont{Tombesi}},\ }%
  \bibfield{journal}{%
  \bibinfo {journal} {Adv.\ At.\ Mol.\ Opt.\ Phys.}\ }%
  \textbf{\bibinfo {volume} {57}},\ \bibinfo {pages} {33} (\bibinfo {year}
  {2009})%
  \bibAnnoteFile{NoStop}{Genes2009}%
\bibitem{Murch2008}%
  \BibitemOpen
  \bibfield{author}{%
  \bibinfo {author} {\bibfnamefont{K.~W.}\ \bibnamefont{Murch}}, \bibinfo
  {author} {\bibfnamefont{K.~L.}\ \bibnamefont{Moore}}, \bibinfo {author}
  {\bibfnamefont{S.}~\bibnamefont{Gupta}},\ and\ \bibinfo {author}
  {\bibfnamefont{D.~M.}\ \bibnamefont{Stamper-Kurn}},\ }%
  \bibfield{journal}{%
  \bibinfo {journal} {Nature Phys.}\ }%
  \textbf{\bibinfo {volume} {4}},\ \bibinfo {pages} {561} (\bibinfo {year}
  {2008})%
  \bibAnnoteFile{NoStop}{Murch2008}%
\bibitem{Brennecke2008}%
  \BibitemOpen
  \bibfield{author}{%
  \bibinfo {author} {\bibfnamefont{F.}~\bibnamefont{Brennecke}}, \bibinfo
  {author} {\bibfnamefont{S.}~\bibnamefont{Ritter}}, \bibinfo {author}
  {\bibfnamefont{T.}~\bibnamefont{Donner}},\ and\ \bibinfo {author}
  {\bibfnamefont{T.}~\bibnamefont{Esslinger}},\ }%
  \bibfield{journal}{%
  \bibinfo {journal} {Science}\ }%
  \textbf{\bibinfo {volume} {322}},\ \bibinfo {pages} {235} (\bibinfo {year}
  {2008})%
  \bibAnnoteFile{NoStop}{Brennecke2008}%
\bibitem{Wilson-Rae2007}%
  \BibitemOpen
  \bibfield{author}{%
  \bibinfo {author} {\bibfnamefont{I.}~\bibnamefont{Wilson-Rae}}, \bibinfo
  {author} {\bibfnamefont{N.}~\bibnamefont{Nooshi}}, \bibinfo {author}
  {\bibfnamefont{W.}~\bibnamefont{Zwerger}},\ and\ \bibinfo {author}
  {\bibfnamefont{T.~J.}\ \bibnamefont{Kippenberg}},\ }%
  \bibfield{journal}{%
  \bibinfo {journal} {Phys.\ Rev.\ Lett.}\ }%
  \textbf{\bibinfo {volume} {99}},\ \bibinfo {pages} {093901} (\bibinfo {year}
  {2007})%
  \bibAnnoteFile{NoStop}{Wilson-Rae2007}%
\bibitem{Marquardt2007}%
  \BibitemOpen
  \bibfield{author}{%
  \bibinfo {author} {\bibfnamefont{F.}~\bibnamefont{Marquardt}}, \bibinfo
  {author} {\bibfnamefont{J.~P.}\ \bibnamefont{Chen}}, \bibinfo {author}
  {\bibfnamefont{A.~A.}\ \bibnamefont{Clerk}},\ and\ \bibinfo {author}
  {\bibfnamefont{S.~M.}\ \bibnamefont{Girvin}},\ }%
  \bibfield{journal}{%
  \bibinfo {journal} {Phys.\ Rev.\ Lett.}\ }%
  \textbf{\bibinfo {volume} {99}},\ \bibinfo {pages} {093902} (\bibinfo {year}
  {2007})%
  \bibAnnoteFile{NoStop}{Marquardt2007}%
\bibitem{Genes2008}%
  \BibitemOpen
  \bibfield{author}{%
  \bibinfo {author} {\bibfnamefont{C.}~\bibnamefont{Genes}}, \bibinfo {author}
  {\bibfnamefont{D.}~\bibnamefont{Vitali}}, \bibinfo {author}
  {\bibfnamefont{P.}~\bibnamefont{Tombesi}}, \bibinfo {author}
  {\bibfnamefont{S.}~\bibnamefont{Gigan}},\ and\ \bibinfo {author}
  {\bibfnamefont{M.}~\bibnamefont{Aspelmeyer}},\ }%
  \bibfield{journal}{%
  \bibinfo {journal} {Phys.\ Rev.\ A}\ }%
  \textbf{\bibinfo {volume} {77}},\ \bibinfo {pages} {033804} (\bibinfo {year}
  {2008})%
  \bibAnnoteFile{NoStop}{Genes2008}%
\bibitem{Zhang2003}%
  \BibitemOpen
  \bibfield{author}{%
  \bibinfo {author} {\bibfnamefont{J.}~\bibnamefont{Zhang}}, \bibinfo {author}
  {\bibfnamefont{K.}~\bibnamefont{Peng}},\ and\ \bibinfo {author}
  {\bibfnamefont{S.~L.}\ \bibnamefont{Braunstein}},\ }%
  \bibfield{journal}{%
  \bibinfo {journal} {Phys.\ Rev.\ A}\ }%
  \textbf{\bibinfo {volume} {68}},\ \bibinfo {pages} {013808} (\bibinfo {year}
  {2003})%
  \bibAnnoteFile{NoStop}{Zhang2003}%
\bibitem{Leonhardt1997}%
  \BibitemOpen
  \bibfield{author}{%
  \bibinfo {author} {\bibfnamefont{U.}~\bibnamefont{Leonhardt}},\ }%
  \emph{\bibinfo {title} {Measuring the Quantum State of Light}}\ (\bibinfo
  {publisher} {Cambridge University Press},\ \bibinfo {year} {1997})%
  \bibAnnoteFile{NoStop}{Leonhardt1997}%
\bibitem{Vitali2007}%
  \BibitemOpen
  \bibfield{author}{%
  \bibinfo {author} {\bibfnamefont{D.}~\bibnamefont{Vitali}}, \bibinfo {author}
  {\bibfnamefont{S.}~\bibnamefont{Gigan}}, \bibinfo {author}
  {\bibfnamefont{A.}~\bibnamefont{Ferreira}}, \bibinfo {author}
  {\bibfnamefont{H.~R.}\ \bibnamefont{B\"ohm}}, \bibinfo {author}
  {\bibfnamefont{P.}~\bibnamefont{Tombesi}}, \bibinfo {author}
  {\bibfnamefont{A.}~\bibnamefont{Guerreiro}}, \bibinfo {author}
  {\bibfnamefont{V.}~\bibnamefont{Vedral}}, \bibinfo {author}
  {\bibfnamefont{A.}~\bibnamefont{Zeilinger}},\ and\ \bibinfo {author}
  {\bibfnamefont{M.}~\bibnamefont{Aspelmeyer}},\ }%
  \bibfield{journal}{%
  \bibinfo {journal} {Phys.\ Rev.\ Lett.}\ }%
  \textbf{\bibinfo {volume} {98}},\ \bibinfo {pages} {030405} (\bibinfo {year}
  {2007})%
  \bibAnnoteFile{NoStop}{Vitali2007}%
\bibitem{Wu1987}%
  \BibitemOpen
  \bibfield{author}{%
  \bibinfo {author} {\bibfnamefont{L.-A.}\ \bibnamefont{Wu}}, \bibinfo {author}
  {\bibfnamefont{M.}~\bibnamefont{Xiao}},\ and\ \bibinfo {author}
  {\bibfnamefont{H.~J.}\ \bibnamefont{Kimble}},\ }%
  \bibfield{journal}{%
  \bibinfo {journal} {J.\ Opt.\ Soc.\ Am.\ B}\ }%
  \textbf{\bibinfo {volume} {4}},\ \bibinfo {pages} {1465} (\bibinfo {year}
  {1987})%
  \bibAnnoteFile{NoStop}{Wu1987}%
\bibitem{Parkins1999}%
  \BibitemOpen
  \bibfield{author}{%
  \bibinfo {author} {\bibfnamefont{A.~S.}\ \bibnamefont{Parkins}}\ and\
  \bibinfo {author} {\bibfnamefont{H.~J.}\ \bibnamefont{Kimble}},\ }%
  \bibfield{journal}{%
  \bibinfo {journal} {J.\ Opt.\ B: Quantum Semiclassical Opt.}\ }%
  \textbf{\bibinfo {volume} {1}},\ \bibinfo {pages} {496} (\bibinfo {year}
  {1999})%
  \bibAnnoteFile{NoStop}{Parkins1999}%
\bibitem{Lukin2000}%
  \BibitemOpen
  \bibfield{author}{%
  \bibinfo {author} {\bibfnamefont{M.~D.}\ \bibnamefont{Lukin}}, \bibinfo
  {author} {\bibfnamefont{S.~F.}\ \bibnamefont{Yelin}},\ and\ \bibinfo {author}
  {\bibfnamefont{M.}~\bibnamefont{Fleischhauer}},\ }%
  \bibfield{journal}{%
  \bibinfo {journal} {Phys.\ Rev.\ Lett.}\ }%
  \textbf{\bibinfo {volume} {84}},\ \bibinfo {pages} {4232} (\bibinfo {year}
  {2000})%
  \bibAnnoteFile{NoStop}{Lukin2000}%
\bibitem{Julsgaard2004}%
  \BibitemOpen
  \bibfield{author}{%
  \bibinfo {author} {\bibfnamefont{B.}~\bibnamefont{Julsgaard}}, \bibinfo
  {author} {\bibfnamefont{J.}~\bibnamefont{Sherson}}, \bibinfo {author}
  {\bibfnamefont{J.~I.}\ \bibnamefont{Cirac}}, \bibinfo {author}
  {\bibfnamefont{J.}~\bibnamefont{Fiur\'{a}\v{s}ek}},\ and\ \bibinfo {author}
  {\bibfnamefont{E.~S.}\ \bibnamefont{Polzik}},\ }%
  \bibfield{journal}{%
  \bibinfo {journal} {Nature}\ }%
  \textbf{\bibinfo {volume} {432}},\ \bibinfo {pages} {482} (\bibinfo {year}
  {2004})%
  \bibAnnoteFile{NoStop}{Julsgaard2004}%
\bibitem{Grangier1998}%
  \BibitemOpen
  \bibfield{author}{%
  \bibinfo {author} {\bibfnamefont{P.}~\bibnamefont{Grangier}}, \bibinfo
  {author} {\bibfnamefont{J.~A.}\ \bibnamefont{Levenson}},\ and\ \bibinfo
  {author} {\bibfnamefont{J.-P.}\ \bibnamefont{Poizat}},\ }%
  \bibfield{journal}{%
  \bibinfo {journal} {Nature}\ }%
  \textbf{\bibinfo {volume} {396}},\ \bibinfo {pages} {537} (\bibinfo {year}
  {1998})%
  \bibAnnoteFile{NoStop}{Grangier1998}%
\bibitem{Verlot2009}%
  \BibitemOpen
  \bibfield{author}{%
  \bibinfo {author} {\bibfnamefont{P.}~\bibnamefont{Verlot}}, \bibinfo {author}
  {\bibfnamefont{A.}~\bibnamefont{Tavernarakis}}, \bibinfo {author}
  {\bibfnamefont{T.}~\bibnamefont{Briant}}, \bibinfo {author}
  {\bibfnamefont{P.-F.}\ \bibnamefont{Cohadon}},\ and\ \bibinfo {author}
  {\bibfnamefont{A.}~\bibnamefont{Heidmann}},\ }%
  \bibfield{journal}{%
  \bibinfo {journal} {Phys.\ Rev.\ Lett.}\ }%
  \textbf{\bibinfo {volume} {102}},\ \bibinfo {pages} {103601} (\bibinfo {year}
  {2009})%
  \bibAnnoteFile{NoStop}{Verlot2009}%
\bibitem{Schliesser2006}%
  \BibitemOpen
  \bibfield{author}{%
  \bibinfo {author} {\bibfnamefont{A.}~\bibnamefont{Schliesser}}, \bibinfo
  {author} {\bibfnamefont{P.}~\bibnamefont{Del'Haye}}, \bibinfo {author}
  {\bibfnamefont{N.}~\bibnamefont{Nooshi}}, \bibinfo {author}
  {\bibfnamefont{K.~J.}\ \bibnamefont{Vahala}},\ and\ \bibinfo {author}
  {\bibfnamefont{T.~J.}\ \bibnamefont{Kippenberg}},\ }%
  \bibfield{journal}{%
  \bibinfo {journal} {Phys.\ Rev.\ Lett.}\ }%
  \textbf{\bibinfo {volume} {97}},\ \bibinfo {pages} {243905} (\bibinfo {year}
  {2006})%
  \bibAnnoteFile{NoStop}{Schliesser2006}%
\bibitem{Corbitt2007}%
  \BibitemOpen
  \bibfield{author}{%
  \bibinfo {author} {\bibfnamefont{T.}~\bibnamefont{Corbitt}}, \bibinfo
  {author} {\bibfnamefont{Y.}~\bibnamefont{Chen}}, \bibinfo {author}
  {\bibfnamefont{E.}~\bibnamefont{Innerhofer}}, \bibinfo {author}
  {\bibfnamefont{H.}~\bibnamefont{M\"uller-Ebhardt}}, \bibinfo {author}
  {\bibfnamefont{D.}~\bibnamefont{Ottaway}}, \bibinfo {author}
  {\bibfnamefont{H.}~\bibnamefont{Rehbein}}, \bibinfo {author}
  {\bibfnamefont{D.}~\bibnamefont{Sigg}}, \bibinfo {author}
  {\bibfnamefont{S.}~\bibnamefont{Whitcomb}}, \bibinfo {author}
  {\bibfnamefont{C.}~\bibnamefont{Wipf}},\ and\ \bibinfo {author}
  {\bibfnamefont{N.}~\bibnamefont{Mavalvala}},\ }%
  \bibfield{journal}{%
  \bibinfo {journal} {Phys.\ Rev.\ Lett.}\ }%
  \textbf{\bibinfo {volume} {98}},\ \bibinfo {pages} {150802} (\bibinfo {year}
  {2007})%
  \bibAnnoteFile{NoStop}{Corbitt2007}%
\bibitem{Teufel2008}%
  \BibitemOpen
  \bibfield{author}{%
  \bibinfo {author} {\bibfnamefont{J.~D.}\ \bibnamefont{Teufel}}, \bibinfo
  {author} {\bibfnamefont{J.~W.}\ \bibnamefont{Harlow}}, \bibinfo {author}
  {\bibfnamefont{C.~A.}\ \bibnamefont{Regal}},\ and\ \bibinfo {author}
  {\bibfnamefont{K.~W.}\ \bibnamefont{Lehnert}},\ }%
  \bibfield{journal}{%
  \bibinfo {journal} {Phys.\ Rev.\ Lett.}\ }%
  \textbf{\bibinfo {volume} {101}},\ \bibinfo {pages} {197203} (\bibinfo {year}
  {2008})%
  \bibAnnoteFile{NoStop}{Teufel2008}%
\bibitem{Wilson2009}%
  \BibitemOpen
  \bibfield{author}{%
  \bibinfo {author} {\bibfnamefont{D.~J.}\ \bibnamefont{Wilson}}, \bibinfo
  {author} {\bibfnamefont{C.~A.}\ \bibnamefont{Regal}}, \bibinfo {author}
  {\bibfnamefont{S.~B.}\ \bibnamefont{Papp}},\ and\ \bibinfo {author}
  {\bibfnamefont{H.~J.}\ \bibnamefont{Kimble}},\ }%
  \bibfield{journal}{%
  \bibinfo {journal} {Phys.\ Rev.\ Lett.}\ }%
  \textbf{\bibinfo {volume} {103}},\ \bibinfo {pages} {207204} (\bibinfo {year}
  {2009})%
  \bibAnnoteFile{NoStop}{Wilson2009}%
\bibitem{Schliesser2008}%
  \BibitemOpen
  \bibfield{author}{%
  \bibinfo {author} {\bibfnamefont{A.}~\bibnamefont{Schliesser}}, \bibinfo
  {author} {\bibfnamefont{R.}~\bibnamefont{Rivi\`{e}re}}, \bibinfo {author}
  {\bibfnamefont{G.}~\bibnamefont{Anetsberger}}, \bibinfo {author}
  {\bibfnamefont{O.}~\bibnamefont{Arcizet}},\ and\ \bibinfo {author}
  {\bibfnamefont{T.~J.}\ \bibnamefont{Kippenberg}},\ }%
  \bibfield{journal}{%
  \bibinfo {journal} {Nature Phys.}\ }%
  \textbf{\bibinfo {volume} {4}},\ \bibinfo {pages} {415} (\bibinfo {year}
  {2008})%
  \bibAnnoteFile{NoStop}{Schliesser2008}%
\bibitem{Leibfried2003}%
  \BibitemOpen
  \bibfield{author}{%
  \bibinfo {author} {\bibfnamefont{D.}~\bibnamefont{Leibfried}}, \bibinfo
  {author} {\bibfnamefont{R.}~\bibnamefont{Blatt}}, \bibinfo {author}
  {\bibfnamefont{C.}~\bibnamefont{Monroe}},\ and\ \bibinfo {author}
  {\bibfnamefont{D.}~\bibnamefont{Wineland}},\ }%
  \bibfield{journal}{%
  \bibinfo {journal} {Rev.\ Mod.\ Phys.}\ }%
  \textbf{\bibinfo {volume} {75}},\ \bibinfo {pages} {281} (\bibinfo {year}
  {2003})%
  \bibAnnoteFile{NoStop}{Leibfried2003}%
\bibitem{Stenholm1986}%
  \BibitemOpen
  \bibfield{author}{%
  \bibinfo {author} {\bibfnamefont{S.}~\bibnamefont{Stenholm}},\ }%
  \bibfield{journal}{%
  \bibinfo {journal} {Rev.\ Mod.\ Phys.}\ }%
  \textbf{\bibinfo {volume} {58}},\ \bibinfo {pages} {699} (\bibinfo {year}
  {1986})%
  \bibAnnoteFile{NoStop}{Stenholm1986}%
\bibitem{Tittonen1999}%
  \BibitemOpen
  \bibfield{author}{%
  \bibinfo {author} {\bibfnamefont{I.}~\bibnamefont{Tittonen}}, \bibinfo
  {author} {\bibfnamefont{G.}~\bibnamefont{Breitenbach}}, \bibinfo {author}
  {\bibfnamefont{T.}~\bibnamefont{Kalkbrenner}}, \bibinfo {author}
  {\bibfnamefont{T.}~\bibnamefont{M\"uller}}, \bibinfo {author}
  {\bibfnamefont{R.}~\bibnamefont{Conradt}}, \bibinfo {author}
  {\bibfnamefont{S.}~\bibnamefont{Schiller}}, \bibinfo {author}
  {\bibfnamefont{E.}~\bibnamefont{Steinsland}}, \bibinfo {author}
  {\bibfnamefont{N.}~\bibnamefont{Blanc}},\ and\ \bibinfo {author}
  {\bibfnamefont{N.~F.}\ \bibnamefont{de~Rooij}},\ }%
  \bibfield{journal}{%
  \bibinfo {journal} {Phys.\ Rev.\ A}\ }%
  \textbf{\bibinfo {volume} {59}},\ \bibinfo {pages} {1038} (\bibinfo {month}
  {Feb}\ \bibinfo {year} {1999})%
  \bibAnnoteFile{NoStop}{Tittonen1999}%
\bibitem{Groeblacher2008}%
  \BibitemOpen
  \bibfield{author}{%
  \bibinfo {author} {\bibfnamefont{S.}~\bibnamefont{Gr\"oblacher}}, \bibinfo
  {author} {\bibfnamefont{S.}~\bibnamefont{Gigan}}, \bibinfo {author}
  {\bibfnamefont{H.~R.}\ \bibnamefont{B\"ohm}}, \bibinfo {author}
  {\bibfnamefont{A.}~\bibnamefont{Zeilinger}},\ and\ \bibinfo {author}
  {\bibfnamefont{M.}~\bibnamefont{Aspelmeyer}},\ }%
  \bibfield{journal}{%
  \bibinfo {journal} {Europhys.\ Lett.}\ }%
  \textbf{\bibinfo {volume} {81}},\ \bibinfo {pages} {54003} (\bibinfo {year}
  {2008})%
  \bibAnnoteFile{NoStop}{Groeblacher2008}%
\bibitem{Diosi2008}%
  \BibitemOpen
  \bibfield{author}{%
  \bibinfo {author} {\bibfnamefont{L.}~\bibnamefont{Di\'{o}si}},\ }%
  \bibfield{journal}{%
  \bibinfo {journal} {Phys.\ Rev.\ A}\ }%
  \textbf{\bibinfo {volume} {78}},\ \bibinfo {pages} {021801(R)} (\bibinfo
  {year} {2008})%
  \bibAnnoteFile{NoStop}{Diosi2008}%
\bibitem{Rabl2009b}%
  \BibitemOpen
  \bibfield{author}{%
  \bibinfo {author} {\bibfnamefont{P.}~\bibnamefont{Rabl}}, \bibinfo {author}
  {\bibfnamefont{C.}~\bibnamefont{Genes}}, \bibinfo {author}
  {\bibfnamefont{K.}~\bibnamefont{Hammerer}},\ and\ \bibinfo {author}
  {\bibfnamefont{M.}~\bibnamefont{Aspelmeyer}},\ }%
  \bibfield{journal}{%
  \bibinfo {journal} {Phys.\ Rev.\ A}\ }%
  \textbf{\bibinfo {volume} {80}},\ \bibinfo {pages} {063819} (\bibinfo {year}
  {2009})%
  \bibAnnoteFile{NoStop}{Rabl2009b}%
\bibitem{Groeblacher2009a}%
  \BibitemOpen
  \bibfield{author}{%
  \bibinfo {author} {\bibfnamefont{S.}~\bibnamefont{Gr\"{o}blacher}}, \bibinfo
  {author} {\bibfnamefont{J.~B.}\ \bibnamefont{Hertzberg}}, \bibinfo {author}
  {\bibfnamefont{M.~R.}\ \bibnamefont{Vanner}}, \bibinfo {author}
  {\bibfnamefont{S.}~\bibnamefont{Gigan}}, \bibinfo {author}
  {\bibfnamefont{K.~C.}\ \bibnamefont{Schwab}},\ and\ \bibinfo {author}
  {\bibfnamefont{M.}~\bibnamefont{Aspelmeyer}},\ }%
  \bibfield{journal}{%
  \bibinfo {journal} {Nature Phys.}\ }%
  \textbf{\bibinfo {volume} {5}},\ \bibinfo {pages} {485} (\bibinfo {year}
  {2009})%
  \bibAnnoteFile{NoStop}{Groeblacher2009a}%
\bibitem{Schliesser2009}%
  \BibitemOpen
  \bibfield{author}{%
  \bibinfo {author} {\bibfnamefont{A.}~\bibnamefont{Schliesser}}, \bibinfo
  {author} {\bibfnamefont{O.}~\bibnamefont{Arcizet}}, \bibinfo {author}
  {\bibfnamefont{R.}~\bibnamefont{Rivi\`{e}re}}, \bibinfo {author}
  {\bibfnamefont{G.}~\bibnamefont{Anetsberger}},\ and\ \bibinfo {author}
  {\bibfnamefont{T.~J.}\ \bibnamefont{Kippenberg}},\ }%
  \bibfield{journal}{%
  \bibinfo {journal} {Nature Phys.}\ }%
  \textbf{\bibinfo {volume} {5}},\ \bibinfo {pages} {509} (\bibinfo {year}
  {2009})%
  \bibAnnoteFile{NoStop}{Schliesser2009}%
\bibitem{Rocheleau2010}%
  \BibitemOpen
  \bibfield{author}{%
  \bibinfo {author} {\bibfnamefont{T.}~\bibnamefont{Rocheleau}}, \bibinfo
  {author} {\bibfnamefont{T.}~\bibnamefont{Ndukum}}, \bibinfo {author}
  {\bibfnamefont{C.}~\bibnamefont{Macklin}}, \bibinfo {author}
  {\bibfnamefont{J.~B.}\ \bibnamefont{Hertzberg}}, \bibinfo {author}
  {\bibfnamefont{A.~A.}\ \bibnamefont{Clerk}},\ and\ \bibinfo {author}
  {\bibfnamefont{K.~C.}\ \bibnamefont{Schwab}},\ }%
  \bibfield{journal}{%
  \bibinfo {journal} {Nature}\ }%
  \textbf{\bibinfo {volume} {463}},\ \bibinfo {pages} {72} (\bibinfo {year}
  {2010})%
  \bibAnnoteFile{NoStop}{Rocheleau2010}%
\bibitem{Cole2008}%
  \BibitemOpen
  \bibfield{author}{%
  \bibinfo {author} {\bibfnamefont{G.~D.}\ \bibnamefont{Cole}}, \bibinfo
  {author} {\bibfnamefont{S.}~\bibnamefont{Gr\"oblacher}}, \bibinfo {author}
  {\bibfnamefont{K.}~\bibnamefont{Gugler}}, \bibinfo {author}
  {\bibfnamefont{S.}~\bibnamefont{Gigan}},\ and\ \bibinfo {author}
  {\bibfnamefont{M.}~\bibnamefont{Aspelmeyer}},\ }%
  \bibfield{journal}{%
  \bibinfo {journal} {Appl.\ Phys.\ Lett.}\ }%
  \textbf{\bibinfo {volume} {92}},\ \bibinfo {pages} {261108} (\bibinfo {year}
  {2008})%
  \bibAnnoteFile{NoStop}{Cole2008}%
\bibitem{Anetsberger2008}%
  \BibitemOpen
  \bibfield{author}{%
  \bibinfo {author} {\bibfnamefont{G.}~\bibnamefont{Anetsberger}}, \bibinfo
  {author} {\bibfnamefont{R.}~\bibnamefont{Rivi\`{e}re}}, \bibinfo {author}
  {\bibfnamefont{A.}~\bibnamefont{Schliesser}}, \bibinfo {author}
  {\bibfnamefont{O.}~\bibnamefont{Arcizet}},\ and\ \bibinfo {author}
  {\bibfnamefont{T.~J.}\ \bibnamefont{Kippenberg}},\ }%
  \bibfield{journal}{%
  \bibinfo {journal} {Nature Photon.}\ }%
  \textbf{\bibinfo {volume} {2}},\ \bibinfo {pages} {627} (\bibinfo {year}
  {2008})%
  \bibAnnoteFile{NoStop}{Anetsberger2008}%
\bibitem{Romero-Isart2010}%
  \BibitemOpen
  \bibfield{author}{%
  \bibinfo {author} {\bibfnamefont{O.}~\bibnamefont{Romero-Isart}}, \bibinfo
  {author} {\bibfnamefont{M.~L.}\ \bibnamefont{Juan}}, \bibinfo {author}
  {\bibfnamefont{R.}~\bibnamefont{Quidant}},\ and\ \bibinfo {author}
  {\bibfnamefont{J.~I.}\ \bibnamefont{Cirac}},\ }%
  \bibfield{journal}{%
  \bibinfo {journal} {New J.\ Phys.}\ }%
  \textbf{\bibinfo {volume} {12}},\ \bibinfo {pages} {033015} (\bibinfo {year}
  {2010})%
  \bibAnnoteFile{NoStop}{Romero-Isart2010}%
\bibitem{Chang2010}%
  \BibitemOpen
  \bibfield{author}{%
  \bibinfo {author} {\bibfnamefont{D.~E.}\ \bibnamefont{Chang}}, \bibinfo
  {author} {\bibfnamefont{C.~A.}\ \bibnamefont{Regal}}, \bibinfo {author}
  {\bibfnamefont{S.~B.}\ \bibnamefont{Papp}}, \bibinfo {author}
  {\bibfnamefont{D.~J.}\ \bibnamefont{Wilson}}, \bibinfo {author}
  {\bibfnamefont{J.}~\bibnamefont{Ye}}, \bibinfo {author}
  {\bibfnamefont{O.}~\bibnamefont{Painter}}, \bibinfo {author}
  {\bibfnamefont{H.~J.}\ \bibnamefont{Kimble}},\ and\ \bibinfo {author}
  {\bibfnamefont{P.}~\bibnamefont{Zoller}},\ }%
  \bibfield{journal}{%
  \bibinfo {journal} {PNAS}\ }%
  \textbf{\bibinfo {volume} {107}},\ \bibinfo {pages} {1005} (\bibinfo {year}
  {2010})%
  \bibAnnoteFile{NoStop}{Chang2010}%
\bibitem{Barker2010}%
  \BibitemOpen
  \bibfield{author}{%
  \bibinfo {author} {\bibfnamefont{P.~F.}\ \bibnamefont{Barker}}\ and\ \bibinfo
  {author} {\bibfnamefont{M.~N.}\ \bibnamefont{Schneider}},\ }%
  \bibfield{journal}{%
  \bibinfo {journal} {Phys.\ Rev.\ A}\ }%
  \textbf{\bibinfo {volume} {81}},\ \bibinfo {pages} {023826} (\bibinfo {year}
  {2010})%
  \bibAnnoteFile{NoStop}{Barker2010}%
\bibitem{Wilson-Rae2004}%
  \BibitemOpen
  \bibfield{author}{%
  \bibinfo {author} {\bibfnamefont{I.}~\bibnamefont{Wilson-Rae}}, \bibinfo
  {author} {\bibfnamefont{P.}~\bibnamefont{Zoller}},\ and\ \bibinfo {author}
  {\bibfnamefont{A.}~\bibnamefont{Imamoglu}},\ }%
  \bibfield{journal}{%
  \bibinfo {journal} {Phys.\ Rev.\ Lett.}\ }%
  \textbf{\bibinfo {volume} {92}},\ \bibinfo {pages} {075507} (\bibinfo {month}
  {02}\ \bibinfo {year} {2004})%
  \bibAnnoteFile{NoStop}{Wilson-Rae2004}%
\bibitem{Poggio2007}%
  \BibitemOpen
  \bibfield{author}{%
  \bibinfo {author} {\bibfnamefont{M.}~\bibnamefont{Poggio}}, \bibinfo {author}
  {\bibfnamefont{C.~L.}\ \bibnamefont{Degen}}, \bibinfo {author}
  {\bibfnamefont{H.~J.}\ \bibnamefont{Mamin}},\ and\ \bibinfo {author}
  {\bibfnamefont{D.}~\bibnamefont{Rugar}},\ }%
  \bibfield{journal}{%
  \bibinfo {journal} {Phys.\ Rev.\ Lett.}\ }%
  \textbf{\bibinfo {volume} {99}},\ \bibinfo {pages} {017201} (\bibinfo {year}
  {2007})%
  \bibAnnoteFile{NoStop}{Poggio2007}%
\bibitem{Kleckner2006b}%
  \BibitemOpen
  \bibfield{author}{%
  \bibinfo {author} {\bibfnamefont{D.}~\bibnamefont{Kleckner}}\ and\ \bibinfo
  {author} {\bibfnamefont{D.}~\bibnamefont{Bouwmeester}},\ }%
  \bibfield{journal}{%
  \bibinfo {journal} {Nature}\ }%
  \textbf{\bibinfo {volume} {444}},\ \bibinfo {pages} {75} (\bibinfo {year}
  {2006})%
  \bibAnnoteFile{NoStop}{Kleckner2006b}%
\bibitem{Corbitt2007b}%
  \BibitemOpen
  \bibfield{author}{%
  \bibinfo {author} {\bibfnamefont{T.}~\bibnamefont{Corbitt}}, \bibinfo
  {author} {\bibfnamefont{C.}~\bibnamefont{Wipf}}, \bibinfo {author}
  {\bibfnamefont{T.}~\bibnamefont{Bodiya}}, \bibinfo {author}
  {\bibfnamefont{D.}~\bibnamefont{Ottaway}}, \bibinfo {author}
  {\bibfnamefont{D.}~\bibnamefont{Sigg}}, \bibinfo {author}
  {\bibfnamefont{N.}~\bibnamefont{Smith}}, \bibinfo {author}
  {\bibfnamefont{S.}~\bibnamefont{Whitcomb}},\ and\ \bibinfo {author}
  {\bibfnamefont{N.}~\bibnamefont{Mavalvala}},\ }%
  \bibfield{journal}{%
  \bibinfo {journal} {Phys.\ Rev.\ Lett.}\ }%
  \textbf{\bibinfo {volume} {99}},\ \bibinfo {pages} {160801} (\bibinfo {year}
  {2007})%
  \bibAnnoteFile{NoStop}{Corbitt2007b}%
\bibitem{Clerk2008}%
  \BibitemOpen
  \bibfield{author}{%
  \bibinfo {author} {\bibfnamefont{A.~A.}\ \bibnamefont{Clerk}}, \bibinfo
  {author} {\bibfnamefont{F.}~\bibnamefont{Marquardt}},\ and\ \bibinfo {author}
  {\bibfnamefont{K.}~\bibnamefont{Jacobs}},\ }%
  \bibfield{journal}{%
  \bibinfo {journal} {New J.\ Phys.}\ }%
  \textbf{\bibinfo {volume} {10}},\ \bibinfo {pages} {095010} (\bibinfo {year}
  {2008})%
  \bibAnnoteFile{NoStop}{Clerk2008}%
\bibitem{Hammerer2009}%
  \BibitemOpen
  \bibfield{author}{%
  \bibinfo {author} {\bibfnamefont{K.}~\bibnamefont{Hammerer}}, \bibinfo
  {author} {\bibfnamefont{M.}~\bibnamefont{Aspelmeyer}}, \bibinfo {author}
  {\bibfnamefont{E.}~\bibnamefont{Polzik}},\ and\ \bibinfo {author}
  {\bibfnamefont{P.}~\bibnamefont{Zoller}},\ }%
  \bibfield{journal}{%
  \bibinfo {journal} {Phys.\ Rev.\ Lett.}\ }%
  \textbf{\bibinfo {volume} {102}},\ \bibinfo {pages} {020501} (\bibinfo {year}
  {2009})%
  \bibAnnoteFile{NoStop}{Hammerer2009}%
\bibitem{Marshall2003}%
  \BibitemOpen
  \bibfield{author}{%
  \bibinfo {author} {\bibfnamefont{W.}~\bibnamefont{Marshall}}, \bibinfo
  {author} {\bibfnamefont{C.}~\bibnamefont{Simon}}, \bibinfo {author}
  {\bibfnamefont{R.}~\bibnamefont{Penrose}},\ and\ \bibinfo {author}
  {\bibfnamefont{D.}~\bibnamefont{Bouwmeester}},\ }%
  \bibfield{journal}{%
  \bibinfo {journal} {Phys.\ Rev.\ Lett.}\ }%
  \textbf{\bibinfo {volume} {91}},\ \bibinfo {pages} {130401} (\bibinfo {year}
  {2003})%
  \bibAnnoteFile{NoStop}{Marshall2003}%
\bibitem{Paternostro2007}%
  \BibitemOpen
  \bibfield{author}{%
  \bibinfo {author} {\bibfnamefont{M.}~\bibnamefont{Paternostro}}, \bibinfo
  {author} {\bibfnamefont{D.}~\bibnamefont{Vitali}}, \bibinfo {author}
  {\bibfnamefont{S.}~\bibnamefont{Gigan}}, \bibinfo {author}
  {\bibfnamefont{M.~S.}\ \bibnamefont{Kim}}, \bibinfo {author}
  {\bibfnamefont{C.}~\bibnamefont{Brukner}}, \bibinfo {author}
  {\bibfnamefont{J.}~\bibnamefont{Eisert}},\ and\ \bibinfo {author}
  {\bibfnamefont{M.}~\bibnamefont{Aspelmeyer}},\ }%
  \bibfield{journal}{%
  \bibinfo {journal} {Phys.\ Rev.\ Lett.}\ }%
  \textbf{\bibinfo {volume} {99}},\ \bibinfo {pages} {250401} (\bibinfo {year}
  {2007})%
  \bibAnnoteFile{NoStop}{Paternostro2007}%
\bibitem{Groeblacher2009b}%
  \BibitemOpen
  \bibfield{author}{%
  \bibinfo {author} {\bibfnamefont{S.}~\bibnamefont{Gr\"oblacher}}, \bibinfo
  {author} {\bibfnamefont{K.}~\bibnamefont{Hammerer}}, \bibinfo {author}
  {\bibfnamefont{M.~R.}\ \bibnamefont{Vanner}},\ and\ \bibinfo {author}
  {\bibfnamefont{M.}~\bibnamefont{Aspelmeyer}},\ }%
  \bibfield{journal}{%
  \bibinfo {journal} {Nature}\ }%
  \textbf{\bibinfo {volume} {460}},\ \bibinfo {pages} {724} (\bibinfo {year}
  {2009})%
  \bibAnnoteFile{NoStop}{Groeblacher2009b}%
\bibitem{Dobrindt2008}%
  \BibitemOpen
  \bibfield{author}{%
  \bibinfo {author} {\bibfnamefont{J.~M.}\ \bibnamefont{Dobrindt}}, \bibinfo
  {author} {\bibfnamefont{I.}~\bibnamefont{Wilson-Rae}},\ and\ \bibinfo
  {author} {\bibfnamefont{T.~J.}\ \bibnamefont{Kippenberg}},\ }%
  \bibfield{journal}{%
  \bibinfo {journal} {Phys.\ Rev.\ Lett.}\ }%
  \textbf{\bibinfo {volume} {101}},\ \bibinfo {pages} {263602} (\bibinfo {year}
  {2008})%
  \bibAnnoteFile{NoStop}{Dobrindt2008}%
\bibitem{Dalibard1985}%
  \BibitemOpen
  \bibfield{author}{%
  \bibinfo {author} {\bibfnamefont{J.}~\bibnamefont{Dalibard}}\ and\ \bibinfo
  {author} {\bibfnamefont{C.}~\bibnamefont{Cohen-Tannoudji}},\ }%
  \bibfield{journal}{%
  \bibinfo {journal} {J.\ Opt.\ Soc.\ Am.\ B}\ }%
  \textbf{\bibinfo {volume} {2}},\ \bibinfo {pages} {1707} (\bibinfo {year}
  {1985})%
  \bibAnnoteFile{NoStop}{Dalibard1985}%
\bibitem{Haroche2006}%
  \BibitemOpen
  \bibfield{author}{%
  \bibinfo {author} {\bibfnamefont{S.}~\bibnamefont{Haroche}}\ and\ \bibinfo
  {author} {\bibfnamefont{J.-M.}\ \bibnamefont{Raimond}},\ }%
  \emph{\bibinfo {title} {Exploring the Quantum: Atoms, Cavities, and
  Photons}}\ (\bibinfo {publisher} {Oxford University Press},\ \bibinfo {year}
  {2006})%
  \bibAnnoteFile{NoStop}{Haroche2006}%
\bibitem{Akram2010}%
  \BibitemOpen
  \bibfield{author}{%
  \bibinfo {author} {\bibfnamefont{U.}~\bibnamefont{Akram}}, \bibinfo {author}
  {\bibfnamefont{N.}~\bibnamefont{Kiesel}}, \bibinfo {author}
  {\bibfnamefont{M.}~\bibnamefont{Aspelmeyer}},\ and\ \bibinfo {author}
  {\bibfnamefont{G.~J.}\ \bibnamefont{Milburn}},\ }%
  \bibfield{journal}{%
  \bibinfo {journal} {arXiv:1002.1517}}%
   (\bibinfo {year} {2010})%
  \bibAnnoteFile{NoStop}{Akram2010}%
\bibitem{Lin2009}%
  \BibitemOpen
  \bibfield{author}{%
  \bibinfo {author} {\bibfnamefont{Q.}~\bibnamefont{Lin}}, \bibinfo {author}
  {\bibfnamefont{J.}~\bibnamefont{Rosenberg}}, \bibinfo {author}
  {\bibfnamefont{X.}~\bibnamefont{Jiang}}, \bibinfo {author}
  {\bibfnamefont{K.~J.}\ \bibnamefont{Vahala}},\ and\ \bibinfo {author}
  {\bibfnamefont{O.}~\bibnamefont{Painter}},\ }%
  \bibfield{journal}{%
  \bibinfo {journal} {Phys.\ Rev.\ Lett.}\ }%
  \textbf{\bibinfo {volume} {103}},\ \bibinfo {pages} {103601} (\bibinfo {year}
  {2009})%
  \bibAnnoteFile{NoStop}{Lin2009}%
\bibitem{Jensen2008}%
  \BibitemOpen
  \bibfield{author}{%
  \bibinfo {author} {\bibfnamefont{K.}~\bibnamefont{Jensen}}, \bibinfo {author}
  {\bibfnamefont{K.}~\bibnamefont{Kim}},\ and\ \bibinfo {author}
  {\bibfnamefont{A.}~\bibnamefont{Zettl}},\ }%
  \bibfield{journal}{%
  \bibinfo {journal} {Nature Nanotech.}\ }%
  \textbf{\bibinfo {volume} {3}},\ \bibinfo {pages} {533} (\bibinfo {year}
  {2008})%
  \bibAnnoteFile{NoStop}{Jensen2008}%
\bibitem{Naik2009}%
  \BibitemOpen
  \bibfield{author}{%
  \bibinfo {author} {\bibfnamefont{A.~K.}\ \bibnamefont{Naik}}, \bibinfo
  {author} {\bibfnamefont{M.~S.}\ \bibnamefont{Hanay}}, \bibinfo {author}
  {\bibfnamefont{W.~K.}\ \bibnamefont{Hiebert}}, \bibinfo {author}
  {\bibfnamefont{X.~L.}\ \bibnamefont{Feng}},\ and\ \bibinfo {author}
  {\bibfnamefont{M.~L.}\ \bibnamefont{Roukes}},\ }%
  \bibfield{journal}{%
  \bibinfo {journal} {Nature Nanotech.}\ }%
  \textbf{\bibinfo {volume} {4}},\ \bibinfo {pages} {445} (\bibinfo {year}
  {2009})%
  \bibAnnoteFile{NoStop}{Naik2009}%
\bibitem{Mamin2001}%
  \BibitemOpen
  \bibfield{author}{%
  \bibinfo {author} {\bibfnamefont{H.~J.}\ \bibnamefont{Mamin}}\ and\ \bibinfo
  {author} {\bibfnamefont{D.}~\bibnamefont{Rugar}},\ }%
  \bibfield{journal}{%
  \bibinfo {journal} {App.\ Phys.\ Lett.}\ }%
  \textbf{\bibinfo {volume} {79}},\ \bibinfo {pages} {3358} (\bibinfo {year}
  {2001})%
  \bibAnnoteFile{NoStop}{Mamin2001}%
\bibitem{Arcizet2006a}%
  \BibitemOpen
  \bibfield{author}{%
  \bibinfo {author} {\bibfnamefont{O.}~\bibnamefont{Arcizet}}, \bibinfo
  {author} {\bibfnamefont{P.-F.}\ \bibnamefont{Cohadon}}, \bibinfo {author}
  {\bibfnamefont{T.}~\bibnamefont{Briant}}, \bibinfo {author}
  {\bibfnamefont{M.}~\bibnamefont{Pinard}}, \bibinfo {author}
  {\bibfnamefont{A.}~\bibnamefont{Heidmann}}, \bibinfo {author}
  {\bibfnamefont{J.-M.}\ \bibnamefont{Mackowski}}, \bibinfo {author}
  {\bibfnamefont{C.}~\bibnamefont{Michel}}, \bibinfo {author}
  {\bibfnamefont{L.}~\bibnamefont{Pinard}}, \bibinfo {author}
  {\bibfnamefont{O.}~\bibnamefont{Fran\c{c}ais}},\ and\ \bibinfo {author}
  {\bibfnamefont{L.}~\bibnamefont{Rousseau}},\ }%
  \bibfield{journal}{%
  \bibinfo {journal} {Phys.\ Rev.\ Lett.}\ }%
  \textbf{\bibinfo {volume} {97}},\ \bibinfo {pages} {133601} (\bibinfo {year}
  {2006})%
  \bibAnnoteFile{NoStop}{Arcizet2006a}%
\bibitem{Munday2009}%
  \BibitemOpen
  \bibfield{author}{%
  \bibinfo {author} {\bibfnamefont{J.~N.}\ \bibnamefont{Munday}}, \bibinfo
  {author} {\bibfnamefont{F.}~\bibnamefont{Capasso}},\ and\ \bibinfo {author}
  {\bibfnamefont{V.~A.}\ \bibnamefont{Parsegian}},\ }%
  \bibfield{journal}{%
  \bibinfo {journal} {Nature}\ }%
  \textbf{\bibinfo {volume} {457}},\ \bibinfo {pages} {170} (\bibinfo {year}
  {2009})%
  \bibAnnoteFile{NoStop}{Munday2009}%
\bibitem{Bleszynski-Jayich2009}%
  \BibitemOpen
  \bibfield{author}{%
  \bibinfo {author} {\bibfnamefont{A.~C.}\ \bibnamefont{Bleszynski-Jayich}},
  \bibinfo {author} {\bibfnamefont{W.~E.}\ \bibnamefont{Shanks}}, \bibinfo
  {author} {\bibfnamefont{B.}~\bibnamefont{Peaudecerf}}, \bibinfo {author}
  {\bibfnamefont{E.}~\bibnamefont{Ginossar}}, \bibinfo {author}
  {\bibfnamefont{F.}~\bibnamefont{von Oppen}}, \bibinfo {author}
  {\bibfnamefont{L.}~\bibnamefont{Glazman}},\ and\ \bibinfo {author}
  {\bibfnamefont{J.~G.~E.}\ \bibnamefont{Harris}},\ }%
  \bibfield{journal}{%
  \bibinfo {journal} {Science}\ }%
  \textbf{\bibinfo {volume} {326}},\ \bibinfo {pages} {272} (\bibinfo {year}
  {2009})%
  \bibAnnoteFile{NoStop}{Bleszynski-Jayich2009}%
\bibitem{Mari2009}%
  \BibitemOpen
  \bibfield{author}{%
  \bibinfo {author} {\bibfnamefont{A.}~\bibnamefont{Mari}}\ and\ \bibinfo
  {author} {\bibfnamefont{J.}~\bibnamefont{Eisert}},\ }%
  \bibfield{journal}{%
  \bibinfo {journal} {Phys.\ Rev.\ Lett.}\ }%
  \textbf{\bibinfo {volume} {103}},\ \bibinfo {pages} {213603} (\bibinfo {year}
  {2009})%
  \bibAnnoteFile{NoStop}{Mari2009}%
\bibitem{Jaehne2009}%
  \BibitemOpen
  \bibfield{author}{%
  \bibinfo {author} {\bibfnamefont{K.}~\bibnamefont{J\"ahne}}, \bibinfo
  {author} {\bibfnamefont{C.}~\bibnamefont{Genes}}, \bibinfo {author}
  {\bibfnamefont{K.}~\bibnamefont{Hammerer}}, \bibinfo {author}
  {\bibfnamefont{M.}~\bibnamefont{Wallquist}}, \bibinfo {author}
  {\bibfnamefont{E.~S.}\ \bibnamefont{Polzik}},\ and\ \bibinfo {author}
  {\bibfnamefont{P.}~\bibnamefont{Zoller}},\ }%
  \bibfield{journal}{%
  \bibinfo {journal} {Phys.\ Rev.\ A}\ }%
  \textbf{\bibinfo {volume} {79}},\ \bibinfo {pages} {063819} (\bibinfo {year}
  {2009})%
  \bibAnnoteFile{NoStop}{Jaehne2009}%
\bibitem{Rabl2009c}%
  \BibitemOpen
  \bibfield{author}{%
  \bibinfo {author} {\bibfnamefont{P.}~\bibnamefont{Rabl}}, \bibinfo {author}
  {\bibfnamefont{S.~J.}\ \bibnamefont{Kolkowitz}}, \bibinfo {author}
  {\bibfnamefont{F.~H.}\ \bibnamefont{Koppens}}, \bibinfo {author}
  {\bibfnamefont{J.~G.~E.}\ \bibnamefont{Harris}}, \bibinfo {author}
  {\bibfnamefont{P.}~\bibnamefont{Zoller}},\ and\ \bibinfo {author}
  {\bibfnamefont{M.~D.}\ \bibnamefont{Lukin}},\ }%
  \bibfield{journal}{%
  \bibinfo {journal} {arXiv:0908.0316}}%
   (\bibinfo {year} {2009})%
  \bibAnnoteFile{NoStop}{Rabl2009c}%
\bibitem{Cleland2004}%
  \BibitemOpen
  \bibfield{author}{%
  \bibinfo {author} {\bibfnamefont{A.~N.}\ \bibnamefont{Cleland}}\ and\
  \bibinfo {author} {\bibfnamefont{M.~R.}\ \bibnamefont{Geller}},\ }%
  \bibfield{journal}{%
  \bibinfo {journal} {Phys.\ Rev.\ Lett.}\ }%
  \textbf{\bibinfo {volume} {93}},\ \bibinfo {pages} {070501} (\bibinfo {year}
  {2004})%
  \bibAnnoteFile{NoStop}{Cleland2004}%
\bibitem{Hammerer2009b}%
  \BibitemOpen
  \bibfield{author}{%
  \bibinfo {author} {\bibfnamefont{K.}~\bibnamefont{Hammerer}}, \bibinfo
  {author} {\bibfnamefont{M.}~\bibnamefont{Wallquist}}, \bibinfo {author}
  {\bibfnamefont{C.}~\bibnamefont{Genes}}, \bibinfo {author}
  {\bibfnamefont{M.}~\bibnamefont{Ludwig}}, \bibinfo {author}
  {\bibfnamefont{F.}~\bibnamefont{Marquardt}}, \bibinfo {author}
  {\bibfnamefont{P.}~\bibnamefont{Treutlein}}, \bibinfo {author}
  {\bibfnamefont{P.}~\bibnamefont{Zoller}}, \bibinfo {author}
  {\bibfnamefont{J.}~\bibnamefont{Ye}},\ and\ \bibinfo {author}
  {\bibfnamefont{H.~J.}\ \bibnamefont{Kimble}},\ }%
  \bibfield{journal}{%
  \bibinfo {journal} {Phys.\ Rev.\ Lett.}\ }%
  \textbf{\bibinfo {volume} {103}},\ \bibinfo {pages} {063005} (\bibinfo {year}
  {2009})%
  \bibAnnoteFile{NoStop}{Hammerer2009b}%
\bibitem{Treutlein2007}%
  \BibitemOpen
  \bibfield{author}{%
  \bibinfo {author} {\bibfnamefont{P.}~\bibnamefont{Treutlein}}, \bibinfo
  {author} {\bibfnamefont{D.}~\bibnamefont{Hunger}}, \bibinfo {author}
  {\bibfnamefont{S.}~\bibnamefont{Camerer}}, \bibinfo {author}
  {\bibfnamefont{T.~W.}\ \bibnamefont{H\"ansch}},\ and\ \bibinfo {author}
  {\bibfnamefont{J.}~\bibnamefont{Reichel}},\ }%
  \bibfield{journal}{%
  \bibinfo {journal} {Phys.\ Rev.\ Lett.}\ }%
  \textbf{\bibinfo {volume} {99}},\ \bibinfo {pages} {140403} (\bibinfo {year}
  {2007})%
  \bibAnnoteFile{NoStop}{Treutlein2007}%
\bibitem{Hammerer2010}%
  \BibitemOpen
  \bibfield{author}{%
  \bibinfo {author} {\bibfnamefont{K.}~\bibnamefont{Hammerer}}, \bibinfo
  {author} {\bibfnamefont{K.}~\bibnamefont{Stannigel}}, \bibinfo {author}
  {\bibfnamefont{C.}~\bibnamefont{Genes}}, \bibinfo {author}
  {\bibfnamefont{P.}~\bibnamefont{Zoller}}, \bibinfo {author}
  {\bibfnamefont{P.}~\bibnamefont{Treutlein}}, \bibinfo {author}
  {\bibfnamefont{S.}~\bibnamefont{Camerer}}, \bibinfo {author}
  {\bibfnamefont{D.}~\bibnamefont{Hunger}},\ and\ \bibinfo {author}
  {\bibfnamefont{T.}~\bibnamefont{H\"ansch}},\ }%
  \bibfield{journal}{%
  \bibinfo {journal} {arXiv:1002.4646}}%
   (\bibinfo {year} {2010})%
  \bibAnnoteFile{NoStop}{Hammerer2010}%
\bibitem{Wallquist2010}%
  \BibitemOpen
  \bibfield{author}{%
  \bibinfo {author} {\bibfnamefont{M.}~\bibnamefont{Wallquist}}, \bibinfo
  {author} {\bibfnamefont{K.}~\bibnamefont{Hammerer}}, \bibinfo {author}
  {\bibfnamefont{P.}~\bibnamefont{Zoller}}, \bibinfo {author}
  {\bibfnamefont{C.}~\bibnamefont{Genes}}, \bibinfo {author}
  {\bibfnamefont{M.}~\bibnamefont{Ludwig}}, \bibinfo {author}
  {\bibfnamefont{F.}~\bibnamefont{Marquardt}}, \bibinfo {author}
  {\bibfnamefont{P.}~\bibnamefont{Treutlein}}, \bibinfo {author}
  {\bibfnamefont{J.}~\bibnamefont{Ye}},\ and\ \bibinfo {author}
  {\bibfnamefont{H.~J.}\ \bibnamefont{Kimble}},\ }%
  \bibfield{journal}{%
  \bibinfo {journal} {Phys.\ Rev.\ A}\ }%
  \textbf{\bibinfo {volume} {81}},\ \bibinfo {pages} {023816} (\bibinfo {year}
  {2010})%
  \bibAnnoteFile{NoStop}{Wallquist2010}%
\bibitem{Arndt2009}%
  \BibitemOpen
  \bibfield{author}{%
  \bibinfo {author} {\bibfnamefont{M.}~\bibnamefont{Arndt}}, \bibinfo {author}
  {\bibfnamefont{M.}~\bibnamefont{Aspelmeyer}},\ and\ \bibinfo {author}
  {\bibfnamefont{A.}~\bibnamefont{Zeilinger}},\ }%
  \bibfield{journal}{%
  \bibinfo {journal} {Fortschr.\ Phys.}\ }%
  \textbf{\bibinfo {volume} {57}},\ \bibinfo {pages} {1153} (\bibinfo {year}
  {2009})%
  \bibAnnoteFile{NoStop}{Arndt2009}%
\bibitem{Leggett2002a}%
  \BibitemOpen
  \bibfield{author}{%
  \bibinfo {author} {\bibfnamefont{A.~J.}\ \bibnamefont{Leggett}},\ }%
  \bibfield{journal}{%
  \bibinfo {journal} {J.\ Phys.:\ Condens.\ Matter}\ }%
  \textbf{\bibinfo {volume} {14}},\ \bibinfo {pages} {R415} (\bibinfo {year}
  {2002})%
  \bibAnnoteFile{NoStop}{Leggett2002a}%
\bibitem{Zurek1991}%
  \BibitemOpen
  \bibfield{author}{%
  \bibinfo {author} {\bibfnamefont{W.~H.}\ \bibnamefont{Zurek}},\ }%
  \bibfield{journal}{%
  \bibinfo {journal} {Phys.\ Today}\ }%
  \textbf{\bibinfo {volume} {44}},\ \bibinfo {pages} {36} (\bibinfo {year}
  {1991})%
  \bibAnnoteFile{NoStop}{Zurek1991}%
\bibitem{Penrose2000}%
  \BibitemOpen
  \bibfield{author}{%
  \bibinfo {author} {\bibfnamefont{R.}~\bibnamefont{Penrose}},\ }%
  \enquote{\bibinfo {title} {Quantum [un]speakables: from bell to quantum
  information},}\ in\ \emph{\bibinfo {booktitle} {Quantum [un]speakables: from
  {B}ell to quantum information}},\ \bibinfo {editor} {edited by\ \bibinfo
  {editor} {\bibfnamefont{R.~A.}\ \bibnamefont{Bertlmann}}\ and\ \bibinfo
  {editor} {\bibfnamefont{A.}~\bibnamefont{Zeilinger}}}\ (\bibinfo {publisher}
  {Springer},\ \bibinfo {year} {2000})\ Chap.\ \bibinfo {chapter} {John {B}ell,
  {S}tate {R}eduction and {Q}uanglement}, pp.\ \bibinfo {pages} {319--330}%
  \bibAnnoteFile{NoStop}{Penrose2000}%
\bibitem{Diosi2000}%
  \BibitemOpen
  \bibfield{author}{%
  \bibinfo {author} {\bibfnamefont{L.}~\bibnamefont{Di\'{o}si}},\ }%
  \enquote{\bibinfo {title} {Decoherence: Theoretical, experimental, and
  conceptual problems},}\ \ (\bibinfo {publisher} {Springer},\ \bibinfo {year}
  {2000})\ Chap.\ \bibinfo {chapter} {Emergence of Classicality: From Collapse
  Phenomenologies to Hybrid Dynamics}, pp.\ \bibinfo {pages} {243--250}%
  \bibAnnoteFile{NoStop}{Diosi2000}%
\bibitem{Adler2009}%
  \BibitemOpen
  \bibfield{author}{%
  \bibinfo {author} {\bibfnamefont{S.~L.}\ \bibnamefont{Adler}}\ and\ \bibinfo
  {author} {\bibfnamefont{A.}~\bibnamefont{Bassi}},\ }%
  \bibfield{journal}{%
  \bibinfo {journal} {Science}\ }%
  \textbf{\bibinfo {volume} {325}},\ \bibinfo {pages} {275} (\bibinfo {year}
  {2009})%
  \bibAnnoteFile{NoStop}{Adler2009}%
\end{thebibliography}
\end{document}